\documentclass[aps,twocolumn,showpacs,preprintnumbers,amsmath,amssymb,nofootinbib,showkeys]{revtex4-1}

\RequirePackage[T1]{fontenc}

\usepackage{epsfig,graphicx,amsmath,amssymb}
\RequirePackage{mathptmx}      
\RequirePackage{flushend}
\RequirePackage{color}
\usepackage{changes}

\begin{document}

\title{The $\eta^\prime$ transition form factor from space- and time-like experimental data}

\author{R.~Escribano$^{1,2}$}\email{rescriba@ifae.es}
\author{S.~Gonz\`alez-Sol\'is$^2$}\email{sgonzalez@ifae.cat}
\author{P.~Masjuan$^3$} \email{masjuan@kph.uni-mainz.de}
\author{P.~Sanchez-Puertas$^3$} \email{sanchezp@kph.uni-mainz.de}

\affiliation{$^1$Grup de F\'{\i}sica Te\`orica, Departament de F\'{\i}sica, Universitat Aut\`onoma de Barcelona, 
E-08193 Bellaterra (Barcelona), Spain\\
$^2$Institut de F\'{\i}sica d'Altes Energies (IFAE), The Barcelona Institute of Science and Technology (BIST), Campus UAB, E-08193 Bellaterra (Barcelona), Spain\\
$^3$PRISMA Cluster of Excellence, Institut f\"ur Kernphysik, Johannes Gutenberg-Universit\"at,
D-55099 Mainz, Germany}    

\preprint{MITP/15-105}

\begin{abstract}
The $\eta^\prime$ transition form factor is reanalyzed in view of the recent BESIII first observation of
the Dalitz decay $\eta^\prime\to\gamma e^+e^-$
in both space- and time-like regions at low and intermediate energies using the Pad\'e approximants method. 
The present analysis provides a suitable parameterization for reproducing the measured form factor in the whole energy region
and allows to extract the corresponding low-energy parameters together with a prediction of its values at the origin,
related to $\Gamma_{\eta^\prime\to\gamma\gamma}$, and the asymptotic limit.
The $\eta$-$\eta^\prime$ mixing is reassessed within a mixing scheme compatible with the large-$N_c$ chiral perturbation theory at next-to-leading order, with particular attention to the OZI-rule--violating parameters. The $J/\psi$, $Z\to\eta^{(\prime)}\gamma$ decays are also considered and predictions reported.


\end{abstract}


\maketitle

\section{Introduction}
\label{Intro}
Pad\'e approximants (PAs) have been shown recently to be very useful for describing meson transition form factors
from the analysis of space-like (SL) experimental data
\cite{Masjuan:2012wy,Escribano:2013kba,Sanchez-Puertas:2014zla,Masjuan:2012qn,Masjuan:2014rea}\footnote{See
also the work in Ref.~\cite{Masjuan:2008fv} for the application of PAs to the case of the pion vector form factor.}.
Such parameterizations based on the measurement of SL data have been used to extrapolate our knowledge of the form factors
down to the low-energy limit $(Q^2\to 0)$, thus extracting the low-energy parameters (LEPs),
and up to the high-energy limit $(Q^2\to\infty)$, then predicting the asymptotic behavior.
Moreover, they have been employed to reconstruct the double-virtual transition form factor \cite{Masjuan:2015lca,Escribano:2015vjz}.
PAs are now regarded as a valuable tool for incorporating available data into problems requiring a precise error estimation.
They conform a data-driven approach that can be considered as simple, systematic and model independent,
the latter because one can provide a systematic error which can be reduced as soon as more experimental data is included.
These PAs applied to the pseudoscalar transition form factors (TFFs)
are utilized in the evaluation of the lightest pseudoscalar mesons contributions
to the hadronic light-by-light piece of the anomalous magnetic moment of the muon
\cite{Masjuan:2012wy,Escribano:2013kba,Masjuan:2012qn,Masjuan:2014rea,Sanchez-Puertas:2016mmz},
the calculation of the $\pi^0\to e^+e^-$ \cite{Masjuan:2015lca} and $\eta,\eta' \to \ell^+\ell^-$ rare decays \cite{Masjuan:2015cjl},
the extraction of the $\eta$-$\eta^\prime$ mixing parameters \cite{Escribano:2013kba,Escribano:2015nra},
the analysis of $\pi^{0}$, $\eta$ and $\eta^{\prime}$ single and double Dalitz decays \cite{Escribano:2015vjz},
and in the quest for dark photons \cite{Gardner:2015wea}.
In all cases, they provide an excellent laboratory for synergic studies between theory and experiment.

The PAs $P^L_M(Q^2)$ to a given function $f(Q^2)$ are ratios of two polynomials
(with degree $L$ and $M$, respectively),
constructed such that their Taylor expansion around the origin exactly coincides with that of the function up to the highest possible order,
i.e., $f(Q^2)-P^{L}_{M}(Q^{2})={\mathcal O}(Q^2)^{L+M+1}$ \cite{Baker,Queralt:2010sv}.
They often provide a means of obtaining information about the function outside its circle of convergence,
and more rapidly evaluating the function within it.
However, in spite of being flexible and user friendly, PAs reconstructed from their power series at the origin are rational functions
with a simple analytical structure given by a set of poles.
Therefore, they do not possess branch cuts and cannot be used to predict the position of resonance poles,
which are hidden in the second Riemann sheet of the complex energy plane.
Similarly, PAs reconstructed using information on the branch cut,
which allow for a precise determination of the resonance pole parameters \cite{Masjuan:2013jha,Masjuan:2014psa,Caprini:2016uxy}, 
are not suitable for the extraction of the LEPs, i.e., PAs in their simplest form considered here cannot access different Riemann sheets.
Nonetheless, for special kind of functions, such as Stieltjes \cite{Masjuan:2008cp,Masjuan:2009wy}
or non-Stieltjes but meromorphic functions \cite{Masjuan:2007ay},
convergence theorems for PAs are known.
To apply these theorems, an understanding of the analytical properties of the functions is required in advanced.
When this knowledge is missing, the practitioner would explore a sequence of PAs and expect a pattern of convergence. Even when convergence is guaranteed in advanced, this will be restricted by the limits of the theorem's applicability. The question is whether observing convergence beyond these limits one could infer, within some uncertainties, the approximate analytical structure of the function under consideration.

In this work, we will explore this last insight taking the $\eta^\prime$ TFF as a proof of concept. Within certain approximations, the authors of Ref.~\cite{Stollenwerk:2011zz} proved the TFF to be a Stieltjes function for which the PA convergence is guaranteed in the SL and the time like (TL) below the production threshold~\cite{Baker}. Not only that, but its rate of convergence is also known~\cite{Baker,Masjuan:2008cp,Masjuan:2009wy}. Nevertheless, convergence along the branch cut is not \textit{a priori} ensured by the mathematical theorem. As we will see later, the particularities of the TFF along the branch cut will decide on the success of our approximation.

We will try to learn and extract from the employed sequence of PAs 
details on the analytical properties of this TFF in the energy regime covered by experimental data.
In our previous analyses of the TFFs from SL data, we have always carefully expressed the limits on the range of applicability of PAs
\cite{Masjuan:2012wy,Escribano:2013kba}. 
Initially, PAs could be analytically continued from the SL region to the TL one, but only up to the first singularity,
usually a branch cut in the form of a production threshold.
For instance, in the case of the single Dalitz decay $\pi^0\to e^+e^-\gamma$
PAs can be safely extended into the TL region up to the pion mass since no branch cuts are present.
On the contrary, for the $\eta\to\ell^+\ell^-\gamma$ decays, with $\ell=e,\mu$, 
the presence of the $\pi\pi$ branch cut could in principle limit the application of PAs in the TL region.
However, the $\eta\to e^+e^-\gamma$ decay and its associated TFF in the TL region was recently measured with great accuracy by the
A2 Collaboration~\cite{Aguar-Bartolome:2013vpw}.
The authors compared their measurement with several theoretical predictions, among them ours,
based on SL parameterizations of the TFF in terms of PAs \cite{Escribano:2013kba},
and found that these PAs show the best agreement with data for the full range of $e^+e^-$ invariant masses reached in the experiment.
This nice result challenged our understanding of the PAs method and triggered, for the first time,
a combined analysis of the $\eta$ TFF from both SL and TL experimental data \cite{Escribano:2015nra}.
The reason for that agreement can be understood by the fact that the branch cut in this decay ($\pi\pi$ unitary cut)
is not resonant inside the available phase-space region since the $\rho$ resonance is well beyond the $\eta$ mass.
The PAs will fail for sure at the first pole encountered on the real axis, or, to be more precise,
will start failing at some point near the pole\footnote{The
question is how close the PAs can approach the pole without failing.
A detailed discussion on this issue for the case of $\eta^\prime$ Dalitz decays can be obtained from \cite{Escribano:2015vjz}.}.
In any case, for the $\eta$ TFF, this pole on the real axis is found to be at $\sqrt{s}\simeq 720$ MeV
for the single-pole parameterization used frequently by the experimental collaborations \cite{Escribano:2015nra}.
Therefore, for the $\eta\to\ell^+\ell^-\gamma$ decays the PAs can also be extended into the TL region up to the $\eta$ mass with excellent accuracy.

The case of the $\eta^\prime\to\ell^+\ell^-\gamma$ Dalitz decays is more cumbersome
since the available phase space this time includes the resonant region.
However, the analysis performed in \cite{Escribano:2013kba} on the $\eta^\prime$ TFF using only SL data
revealed that the pole on the real axis for the single-pole parameterization is located at $\sqrt{s}\simeq 830$ MeV.
In order to estimate the region of influence of this pole one can make use of the half-width rule \cite{Masjuan:2012sk}.
In this case, the $\rho$ and $\omega$ resonances are within the phase-space region and the $\phi$ is not far from its end point.
Taking the values of their masses and widths from the PDG \cite{Agashe:2014kda},
the application of this rule gives $M_{\rm eff}\pm\Gamma_{\rm eff}/2=822\pm 58$ MeV \cite{Escribano:2013kba}.
This value of the effective pole is compatible with the result obtained before from the single-pole parameterization,
thus showing that the pole found at $830$ MeV is somewhat a kind of weighted average of the three existing resonance poles.
The range given by the half-width rule above implies that the region of influence of the former pole is from $764$ MeV to $880$ MeV.
Consequently, for the $\eta^\prime$ TFF the PAs can also be used in a safe manner up to around $760$ MeV in the TL region\footnote{In
\cite{Escribano:2015vjz}, we were more conservative and the half-width rule was applied taking only into account the $\rho$ resonance.
As a result, we obtained that the lowest value of the region of influence in that case was around $700$ MeV.}.

\begin{figure}
\centering
\includegraphics[width=\columnwidth]{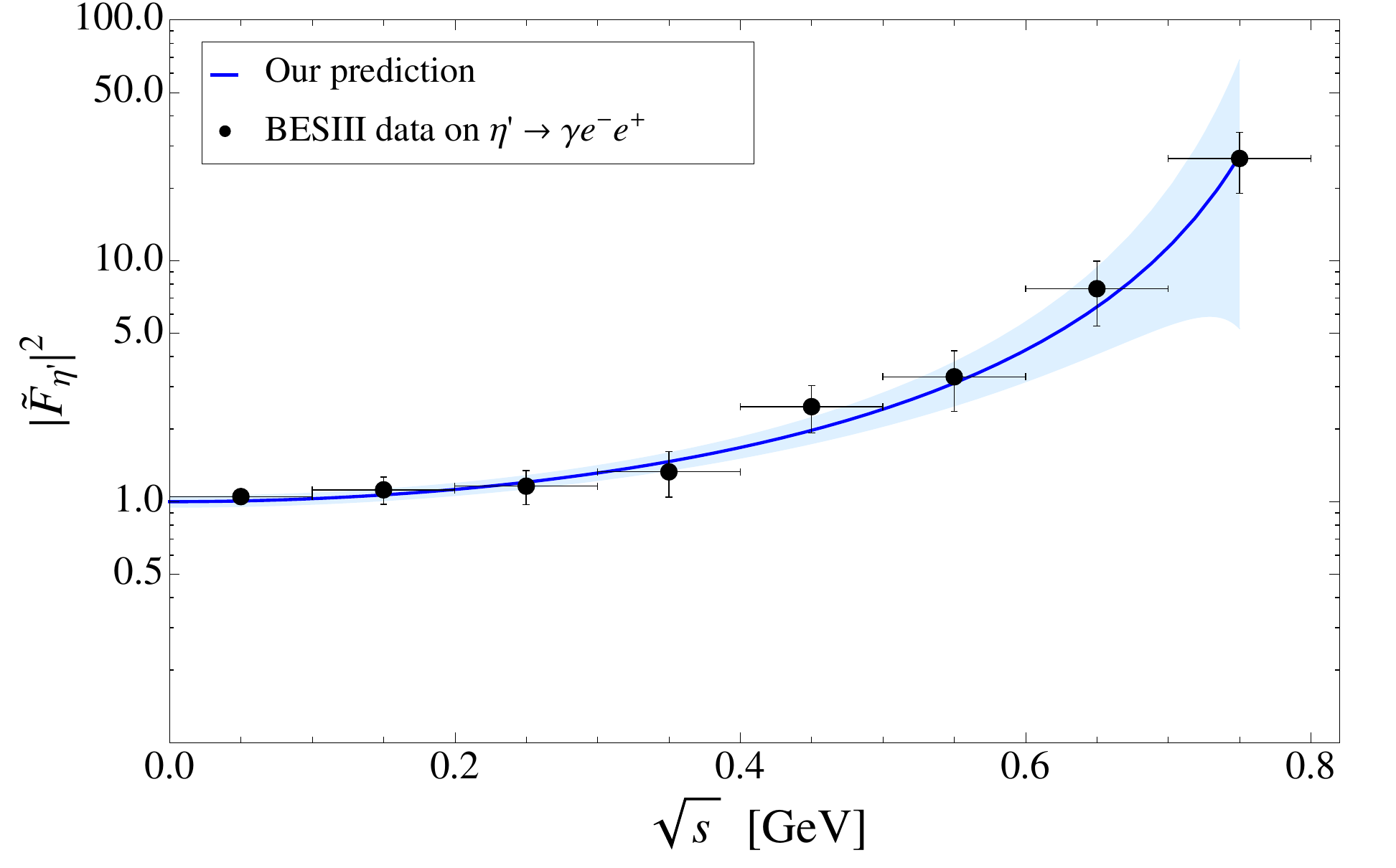}
\caption{Our prediction for the $\eta^{\prime}$ transition form factor in the time-like region
obtained from the $P_{1}^{6}(\sqrt{s})$ fit to space-like data performed in \cite{Escribano:2013kba}.
Experimental points are from the BESIII measurement in \cite{Ablikim:2015wnx}.}
\label{Fig1}
\end{figure}

Recently, the BESIII Collaboration reported the first measurement of the $e^+e^-$ invariant-mass distribution for the
$\eta^\prime\to e^+e^-\gamma$ decay up to $750$ MeV \cite{Ablikim:2015wnx}.
As discussed, our prediction for the TL region of the $\eta^\prime$ TFF based solely on SL data should be able to describe 
this new measurement.
In Figure \ref{Fig1}, the BESIII experimental extraction of the modulus square of the $\eta^\prime$ TFF
as a function of the $e^+e^-$ invariant-mass $(\sqrt{s})$ is compared with our theoretical prediction.
It is worth remarking that this is not a fit but a prediction and the agreement is seen to be excellent.

The main purpose of the present work is therefore to further improve our determination of the $\eta^\prime$ TFF
taking into consideration not only the existing SL experimental data but also the new set of TL data from the recent BESIII measurement.
This combined analysis will allow us to better determine the LEPs of the TFF, its normalization and the asymptotic limit.
Such an enhancement permits to reconsider the $\eta$-$\eta^\prime$ mixing,
with special emphasis on the OZI-rule--violating parameters and the $J/\psi(Z)\to\eta^{(\prime)}\gamma$ decays.
In Section \ref{TLregion}, we comment on the reasons we believe justify  the success of PAs when applied to the TL region.
In Section \ref{TLdata}, we include the TL data in the analysis, present the new results and comment the improvements achieved.
Section \ref{mixing} is devoted to the reassessment of the $\eta$-$\eta^\prime$ mixing parameters within a mixing scheme compatible with the large-$N_c$ chiral perturbation theory at next-to-leading order, with particular attention to the OZI-rule--violating parameters. The consequences of these results for the $J/\psi$ and $Z$ radiative decays are also investigated.
Finally, in Section \ref{conclusions}, we conclude and mention the future prospects.

\begin{figure*}
\centering
\includegraphics[width=\columnwidth]{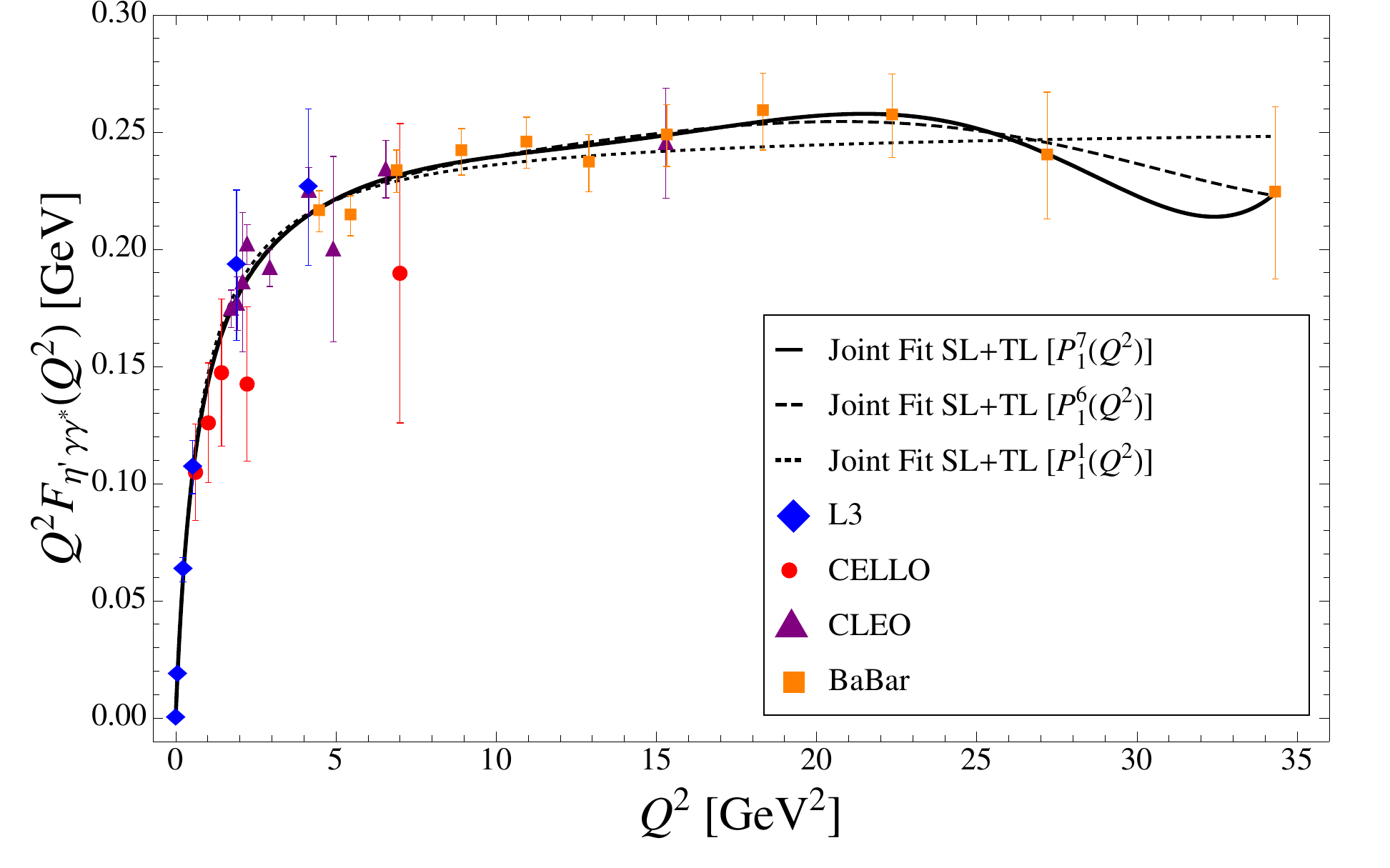}
\includegraphics[width=\columnwidth]{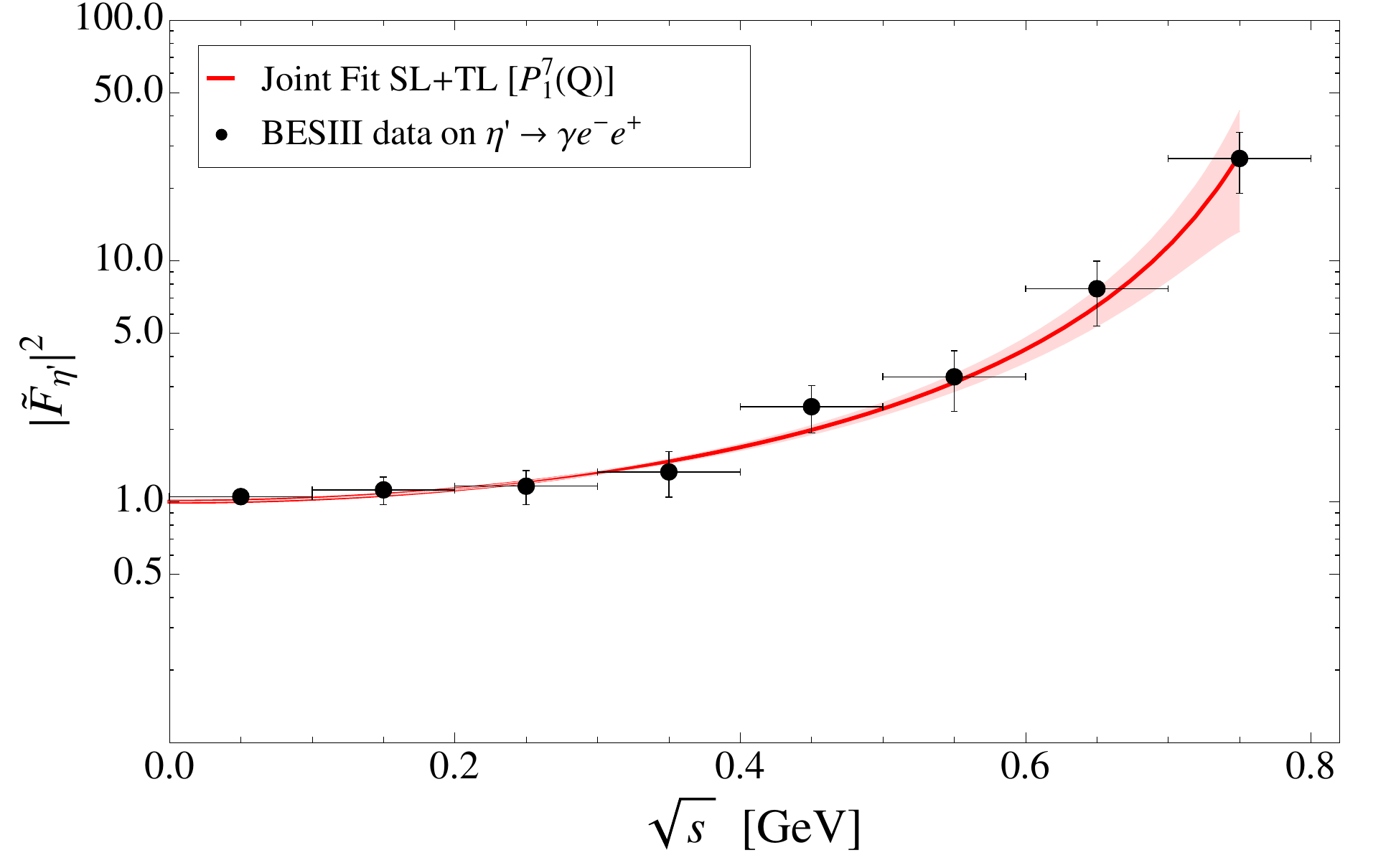}
\caption{$\eta^\prime$ TFF in the SL (left) and TL (right) regions after a joint fit to the SL and TL sets of experimental data.
The dotted, dashed and solid black lines represent the fits using $P_{1}^{1}(Q^{2}), P_{1}^{6}(Q^{2})$ and $P_{1}^{7}(Q^{2})$,
respectively, on the left panel, while the red solid line represents $\widetilde{P}_{1}^{7}(\sqrt{s})$ on the right panel.}
\label{Fitstodata}
\end{figure*}

\section{Pad\'e approximants in the time-like region}
\label{TLregion}

Within certain approximations, the authors of Ref.~\cite{Stollenwerk:2011zz} proved the isovector contribution to the TFF to be a Stieltjes function, which is, that this contribution to the TFF can be represented by an integral form defined as~\cite{Baker}
\begin{equation}\label{Stieltjes}
f(q^2) = \int_0^{1/R} \frac{d \phi(u)}{1-u q^2}\,
\end{equation}
where $\phi(u)$ is any bounded and nondecreasing function~\cite{Baker}. By defining $R=4m_{\pi}^2$, identifying 
$d \phi(u) = {\rm const.}\times\frac{q^2}{\pi}\frac{ {\rm Im}F(1/u)}{u}$, and making the change of variables $u=1/s$, Eq.~(\ref{Stieltjes}) returns the once-subracted dispersive representation of the isovector contribution discussed in Ref.~\cite{Stollenwerk:2011zz}, and also exploited in Refs.~\cite{Hanhart:2013vba,Kubis:2015sga}. Since Im$F(s)= \sigma^3(s) P(s) |F_V(s)|^2$~\cite{Stollenwerk:2011zz} where $\sigma(s)=\sqrt{1-4m_{\pi}^2/s}$, $P(s)$ a linear polynomial with positive slope and $F_V(s)$ the pion vector FF, then  Im$F(s)$  is a positive function, the requirement of $\phi(u)$ to be nondecreasing is fulfilled and the convergence of PAs to the TFF is guaranteed.\footnote{If the function $f(z)$ is a Stieltjes function, its $n^{th}$-subtracted version is a Stieltjes function as well~\cite{Baker}.}

Pad\'e Theory not only provides a convergence theorem for a sequence of PAs to Stieltjes functions, i.e.,  $\lim_{N,M\to \infty} P^N_M(s) - f(s) =0$, but also its rate of convergence~\cite{Baker,Masjuan:2008cp,Masjuan:2009wy}, given by the difference of two consecutive elements in the PA sequence. As we will see later, this error prescription will return very small theoretical uncertainties. To be more conservative, we designed in Refs.~\cite{Masjuan:2008fv,Masjuan:2012wy,Escribano:2013kba,Escribano:2015nra} a different method to extract such uncertainty which yields errors at the level of the statistical ones.

Still, even though the $\pi\pi$ unitary cut driving the decay is of Stieltjes nature, there is no \emph{a priori} reason
why the PA should work above the branch cut.
The cumbersome situation is, however, that at least the $P^L_1(s)$ sequence does work well above the cut (cf.~Figure \ref{Fig1}).
And the unanswered question is, then, whether one could have anticipated this success and how general is for any arbitrary situation.
A fair statement would be to say that, approximately,
the TFF is a meromorphic function which has nothing but a set of single and isolated poles within the data range.
In this scenario, PA are an excellent approximation tool~\cite{Masjuan:2007ay}.
Moreover, they tell us about the underlying physical phenomena driving the decay without the need to assume any model. 

To better understand this situation from a qualitative point of view, let us discuss the following.
As we have said, in the zero-width approximation, the TFF becomes a meromorphic function.
If the TFF contains a single and isolated pole, the $P^L_1(s)$ sequence reproduces the pole of the TFF with infinite precision.
As soon as the width is again switched on, the $\pi \pi$ threshold opens a branch cut responsible for that width.
Then, at first, no mathematical theorem will guarantee convergence on this scenario.
On the contrary, if the convergence theorem is to be satisfied, one would expect the single pole of the $P^L_1(s)$
to be located closer and closer to the threshold point as soon as $L \to \infty$, since this is the first singular point the PA is going to find.

A closer inspection at the threshold expansion of the TFF reveals a different pattern for real and imaginary parts
in terms of the variable $q^2$, the center-of-mass momentum in the $\pi\pi$ rest frame. At low energies, one can model the $P \to \gamma^* \gamma$  TFF as a convolution between the $P \to \pi \pi \gamma$ amplitude with the $\pi \pi \to \gamma^*$ pion electromagnetic form factor $F_V(s)$~\cite{Hanhart:2013vba}. 

The behavior of the $\pi \pi$ branch cut at threshold has then, two different components, both of them well known.
The knowledge of the $\pi \pi$ threshold for $F_V(s)$ comes from the P-wave $\pi\pi$ scattering amplitude (the opened cut yields vector states)
together with the Fermi-Watson theorem that relates the phase of the scattering amplitude with the phase of the form factor below
the first inelastic threshold.
The $\pi\pi$ P-wave scattering amplitude $t^1_1(s)$ at threshold behaves like~\cite{Ananthanarayan:2000ht}:
\begin{eqnarray}\label{t11exp}
\textrm{Im}[t^1_1(s)]&=&q^4\sqrt{q^2}\left(\frac{a^2}{m_\pi} + \frac{4 a\, b\, m_\pi^2 -a^2}{2 m_\pi^3} q^2 + {\cal O}(q^2)^{3} \right)\ ,
\nonumber\\
\textrm{Re}[t^1_1(s)]&=&q^2 \left(a + b\, q^2 + {\cal O}(q^2)^2\right)\ ,
\end{eqnarray}
with $4 q^2 = s-4 m_\pi^2$, and where for the imaginary part we used the unitary relation
Im$[t^1_1(s)^{-1}]=-\sigma(s)$. 
The absolute value of the threshold expansion of the amplitude $t^1_1(s)$ is basically a polynomial in $(s-4m_\pi^2)$ with the influence of its imaginary part starting only at $(s- 4m_\pi^2)^4$.
Following the previous equation, if the threshold parameters $a$ and $b$ are of order $1$
(with the appropriate units)~\cite{Ananthanarayan:2000ht},
then the real part dominates near threshold and the absolute value is given basically by the real part. 
By virtue of the unitary relation for the $F_V(s)$, $\textrm{Im} F_V(s) = \sigma(s)F_V(s) t^1_1(s)^*$, and the expansion in Eq.~(\ref{t11exp}), one concludes that while the real part of the threshold expansion of the $F_V(s)$ starts at order $(s- 4m_\pi^2)^0$, its imaginary part coming from $\sigma(s) \textrm{Re}[t^1_1(s)]$ only starts showing up at order $(s- 4m_\pi^2)^{3/2}$.

In summary, since Im[TFF] $\sim \sigma(s)^3 |F_V(s)|^2$ and $F_V(s)$ at threshold is basically real, near the $\pi\pi$ threshold  the imaginary part of the TFF, and thereby the expected discontinuity, will behave as $(s-4 m_{\pi}^2)^{3/2}$, while its real part as $(s-4 m_{\pi}^2)^{0}$. If this is the case and the offset of the threshold is that smooth, the $P^L_1(s)$ sequence will be an excellent tool to reproduce
the TFF near and above the threshold. Actually, taking the definition of a $P^L_1(s)$ given by
\begin{equation}
\label{PAeq}
P_1^L(s)=\sum_{k=0}^{L-1}a_k s^k+\frac{a_L s^L}{1-\frac{a_{L+1}}{a_L}s}\ ,
\end{equation}
the polynomial part will reproduce the modulus of the $\pi\pi$ discontinuity, and the PA pole part will account in an effective manner for the pole of the TFF far away from the threshold.

The last question is, then, up to what energy one can go above the threshold before failing.
The threshold expansion itself must fail at some point because it breaks unitarity by powers of $(s-4m_\pi^2)$.
A quantitative answer to this question would demand to study this problem using a particular model.
To make a general statement, model independent, and qualitative,
we notice that the threshold expansion should break down when the presence of the resonance pole is large enough and
cannot be approximated by a polynomial in $(s-4m_\pi^2)$.
This happens basically at a distance of the pole given by the half-width rule~\cite{Masjuan:2012sk} which,
as argued in the Introduction, provides a simple estimate of the PA range.

The previous discussion already excludes the generalization of our results for any arbitrary Stieltjes function since
the clue feature is the behavior around the threshold point. While for vector and tensor form factors we foresee good performance of our PA method, for a scalar form factor with an abrupt threshold offset, the range of applicability within the time-like region will be more limited.
Parameterizations existent in literature which would be suitable to be compared our method with can be found in~\cite{Landsberg:1986fd,Ametller:1991jv,Feldmann:1998yc,
Balakireva:2011wp,Czyz:2012nq,Geng:2012qg,Klopot:2012hd,Klopot:2013laa,Hanhart:2013vba,Roig:2014uja,Agaev:2014wna,Kubis:2015sga}.

\begin{table}
\centering
\begin{tabular}{lcccccc}
\hline
\multicolumn{1}{c}{}& 
\multicolumn{2}{c}{\bf{Constraining} $F_{\eta^{\prime}\gamma\gamma}(0)$}& 
\multicolumn{2}{c}{\bf{Predicting} $F_{\eta^{\prime}\gamma\gamma}(0)$}\\[1ex]
&\textbf{$P_{1}^{7}$}& \textbf{$P_{1}^{1}$}&\textbf{$P_{1}^{6}$}& \textbf{$P_{1}^{1}$}\\
\hline
$b_{\eta^{\prime}}$&1.31(4)&1.25(3)&1.30(4)&1.27(4)\\[0.5ex]
$c_{\eta^{\prime}}$&1.74(9)&1.56(6)&1.73(9)&1.62(11)\\[0.5ex]
$d_{\eta^{\prime}}$&2.30(19)&1.94(12)&2.29(19)&2.06(22)\\[0.5ex]
$F_{\eta^{\prime}\gamma\gamma}(0)$&0.344(5)&0.345(5)&0.342(13)&0.351(10)\\[0.5ex]
$Q^{2}F_{\eta^{\prime}\gamma^{*}\gamma}^{\rm asym}(Q^{2})$&
$\rule{5mm}{0.1mm}$&0.254(3)&$\rule{5mm}{0.1mm}$&0.253(3)\\[0.5ex]
$\sqrt{s_{p}}$ (GeV)&$0.833(14)$&$0.857(9)$&$0.831(13)$&$0.849(15)$\\[0.5ex]
$\chi^{2}_{\rm dof}$&0.65&0.67&0.66&0.68\\[0.5ex]
\hline
\end{tabular}
\caption{Low-energy parameters as obtained after a joint fit to both space- and time-like data with and without including
the measured two-photon partial width as a restriction in the $\chi^{2}$ function of (\ref{chi2}), second and third multicolumn, respectively.
The leading coefficient of the TFF asymptotic limit, the pole of the PA and the $\chi^{2}_{\rm dof}$ are also shown.}
\label{JointFit}
\end{table}

\section{Incorporation of the low-energy time-like data}
\label{TLdata}
Since our goal is to provide a parameterization of the TFF as accurate as possible and we have shown in the previous section
that the TL experimental data up to $0.75$ GeV can be well described with our old parameterization based on SL data,
in this section we will include the TL data as a new data set to be fitted, following~\cite{Escribano:2015nra}. 
At low-momentum transfer, the TFF can be described by the expansion
\begin{eqnarray}
F_{\eta^{\prime}\gamma^{*}\gamma}(Q^{2})&&=F_{\eta^{\prime}\gamma\gamma}(0)\times\nonumber\\&&\times\left(1-b_{\eta^{\prime}}\frac{Q^{2}}{m_{\eta^{\prime}}^{2}}+c_{\eta^{\prime}}\frac{Q^{4}}{m_{\eta^{\prime}}^{4}}-d_{\eta^{\prime}}\frac{Q^{6}}{m_{\eta^{\prime}}^{6}}+\cdots\right)\,,
\end{eqnarray}
where $F_{\eta^{\prime}\gamma\gamma}(0)$ is the normalization (the TFF at zero momentum transfer) while the LEPs parameters $b_{\eta^{\prime}}$, $c_{\eta^{\prime}}$ and $d_{\eta^{\prime}}$ are, respectively, the slope, the curvature and the third derivative of the TFF.
By reassessing our SL fits~\cite{Escribano:2013kba} through including TL data,
we will update the results for the LEPs of the $\eta^{\prime}$ TFF.
The $\chi^{2}$ function minimized in our fit is given by ($\widetilde{F}(\sqrt{s})=F(\sqrt{s})/F(0)$)\footnote{In this work the Pad\'{e} sequences are referred to $Q^{2}F_{\eta^{\prime}\gamma^{*}\gamma}(Q^{2})$.}
\begin{equation}
\begin{array}{l}
\chi^{2}=\sum_{i=1}^{50}\left(
\frac{\left|P^{L}_{M}(Q^{2})\right|_{i}-Q^{2}\left|F^{{\rm{\rm exp}}}_{\eta^{\prime}\gamma^\ast\gamma}(Q^{2})\right|_{i}}
{\sigma_{Q^{2}\left|F^{{\rm{\rm exp}}}_{\eta^{\prime}\gamma^\ast\gamma}(Q^{2})\right|_{i}}}\right)^{2}+\\[4ex]
\sum_{i=1}^{8}\left(
\frac{\left|\widetilde{P}^{L}_{M}(\sqrt{s})\right|_{i}^2-\left|\widetilde{F}^{{\rm{\rm exp}}}_{\eta^{\prime}\gamma^\ast\gamma}(\sqrt{s})\right|_{i}^2}
{\sigma_{\left|\widetilde{F}^{{\rm{\rm exp}}}_{\eta^{\prime}\gamma^\ast\gamma}(\sqrt{s})\right|_{i} ^2}}\right)^{2}+
\left(\frac{P^{L-1}_M(0)-\left|F^{{\rm{\rm exp}}}_{\eta^{\prime}\gamma\gamma}(0)\right|}
{\sigma_{\left|F^{{\rm{\rm exp}}}_{\eta^{\prime}\gamma\gamma}(0)\right|}}\right)^{2}\ ,
\end{array}
\label{chi2}
\end{equation}
where the first and second terms correspond to SL \cite{Behrend:1990sr,Gronberg:1997fj,BABAR:2011ad,Acciarri:1997yx}
and TL \cite{Ablikim:2015wnx} data, respectively,
while the last term encodes information from the TFF at zero momentum transfer and introduces an additional restriction.
For the experimental value we use $F^{{\rm{\rm exp}}}_{\eta^{\prime}\gamma\gamma}(0)=0.3437(55)$ GeV$^{-1}$,
inferred from the partial width to two photons, $\Gamma_{\eta^{\prime}\to\gamma\gamma}=4.35(14)$~keV \cite{Agashe:2014kda},
through the relation
\begin{equation}
\left|F_{\eta^{\prime}\gamma\gamma}(0)\right|^{2}=
\frac{64\pi}{(4\pi\alpha)^{2}}\frac{\Gamma(\eta^{\prime}\rightarrow\gamma\gamma)}{m_{\eta^{\prime}}^{3}}\ .
\label{Pgammagamma0}
\end{equation}
The value $\Gamma_{\eta^{\prime}\to\gamma\gamma}=4.35(14)$~keV cited in~\cite{Agashe:2014kda} is not a measured quantity,
rather a fit inferred from the branching ratio and using the current $\eta^\prime$ total width.
The average experimental determination for such decay reads $4.28(19)$~keV.
It will be interesting to see whether this $0.3\sigma$ difference would affect our results at the precision we are working. 

\begin{figure*}
\centering
\includegraphics[width=\columnwidth]{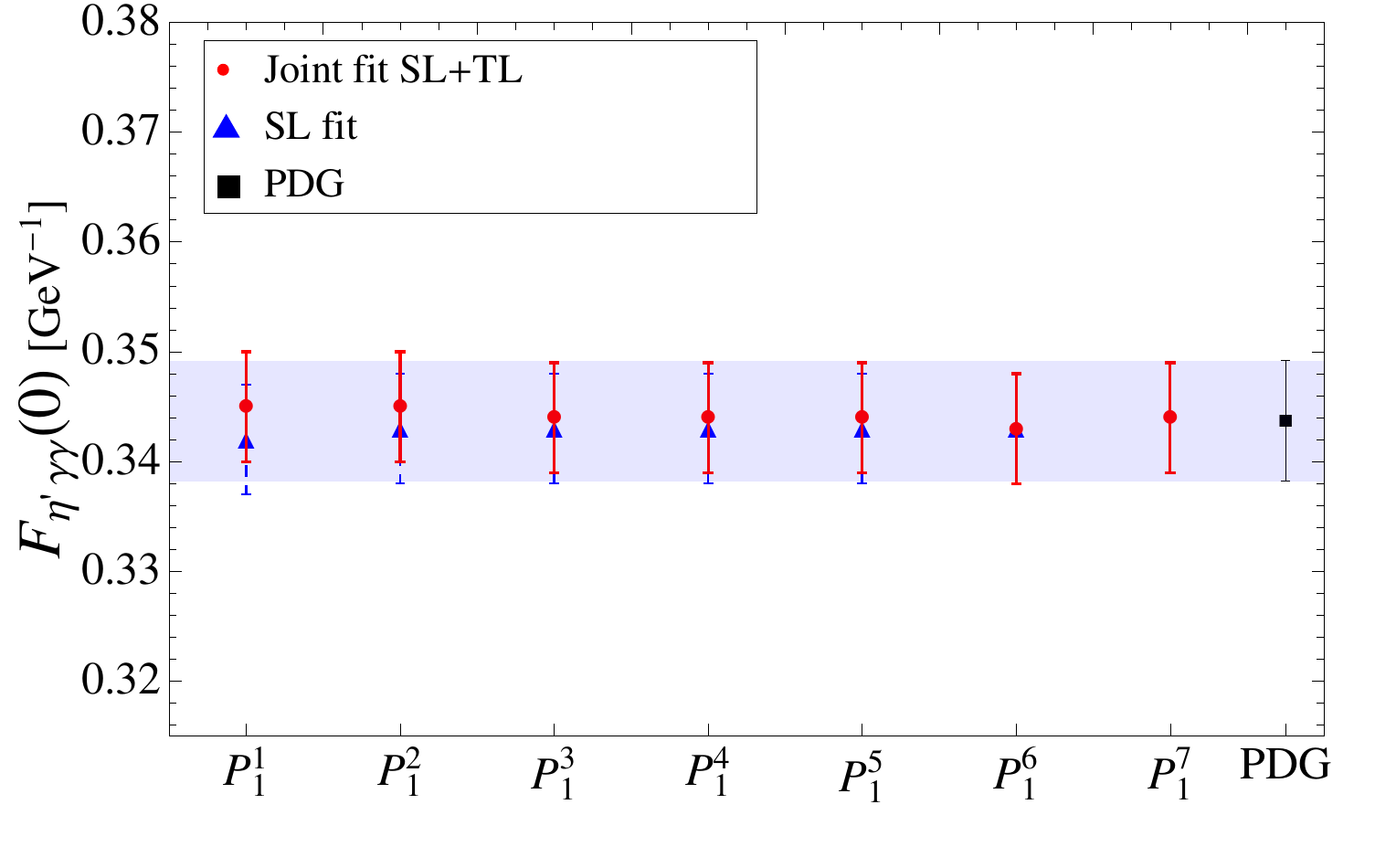}
\includegraphics[width=\columnwidth]{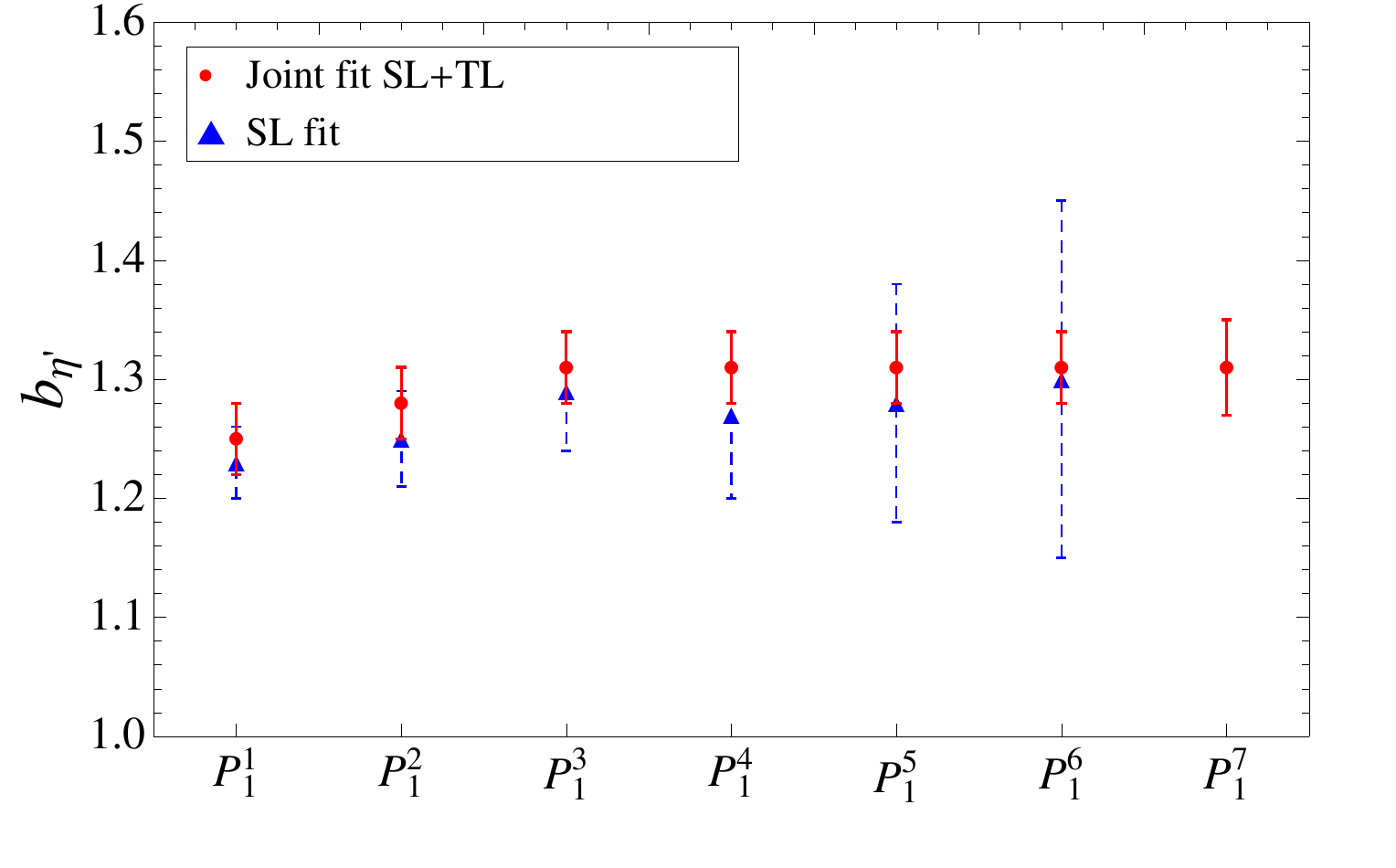}\\[1ex]
\includegraphics[width=\columnwidth]{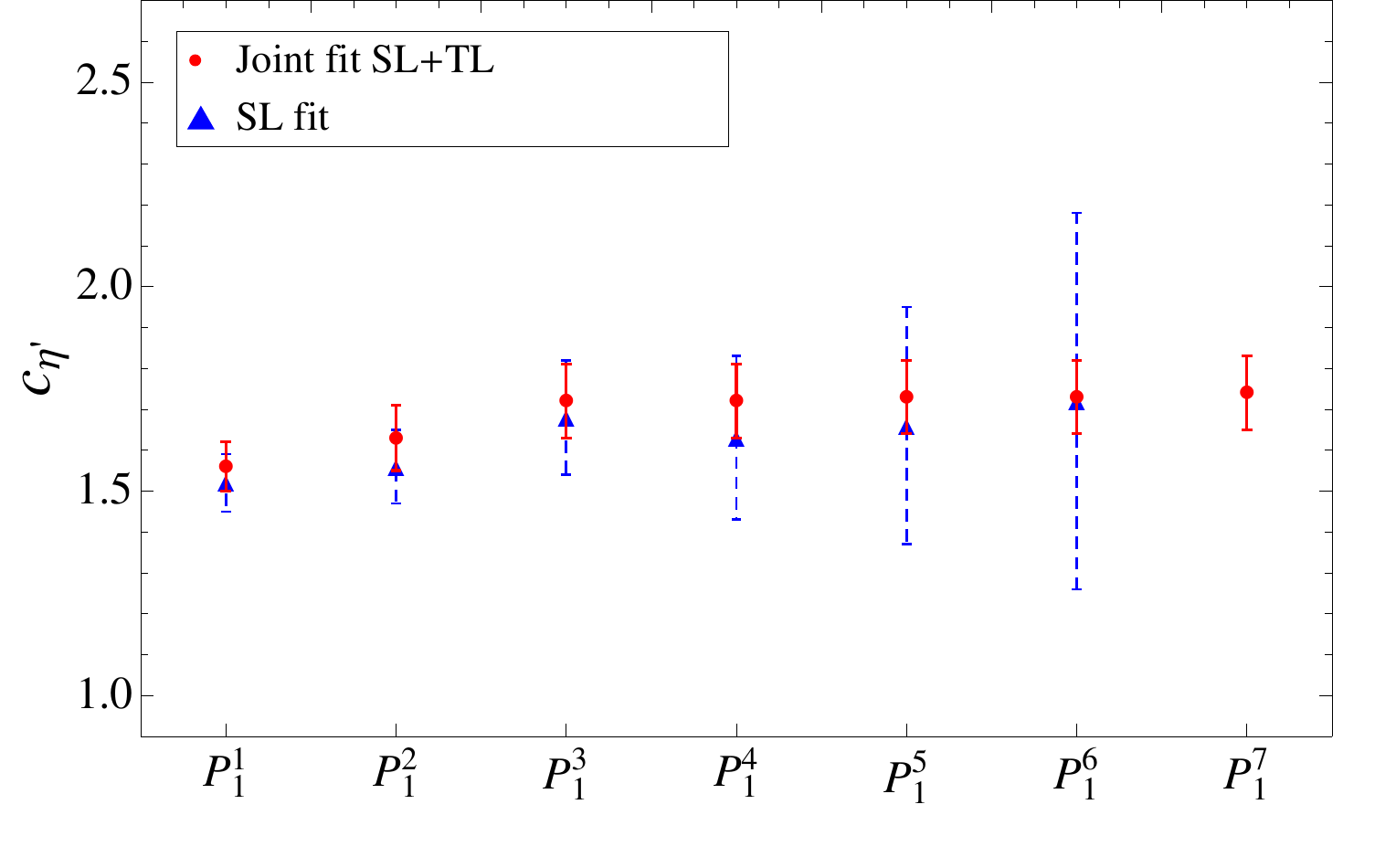}
\includegraphics[width=\columnwidth]{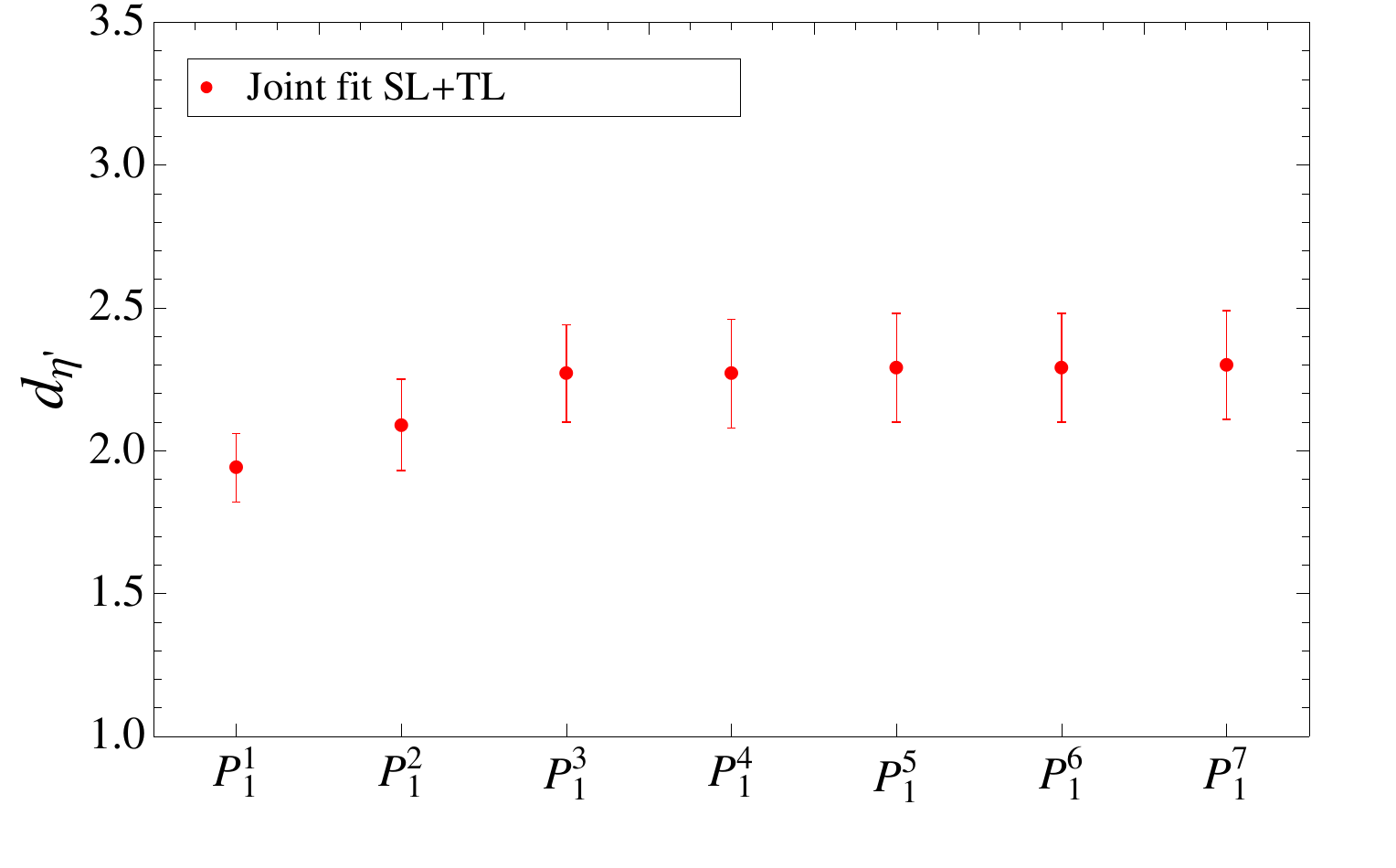}
\caption{Convergence pattern of the $P^L_1$ sequence for
$F_{\eta^{\prime}\gamma\gamma}(0)$, $b_{\eta^{\prime}}$, $c_{\eta^{\prime}}$,
and $d_{\eta^{\prime}}$ as obtained from fitting experimental SL and TL data together with $F_{\eta^{\prime}\gamma\gamma}(0)$
from the PDG~\cite{Agashe:2014kda}.}
\label{Fitpredictionscons}
\end{figure*}

\begin{figure*}
\centering
\includegraphics[width=\columnwidth]{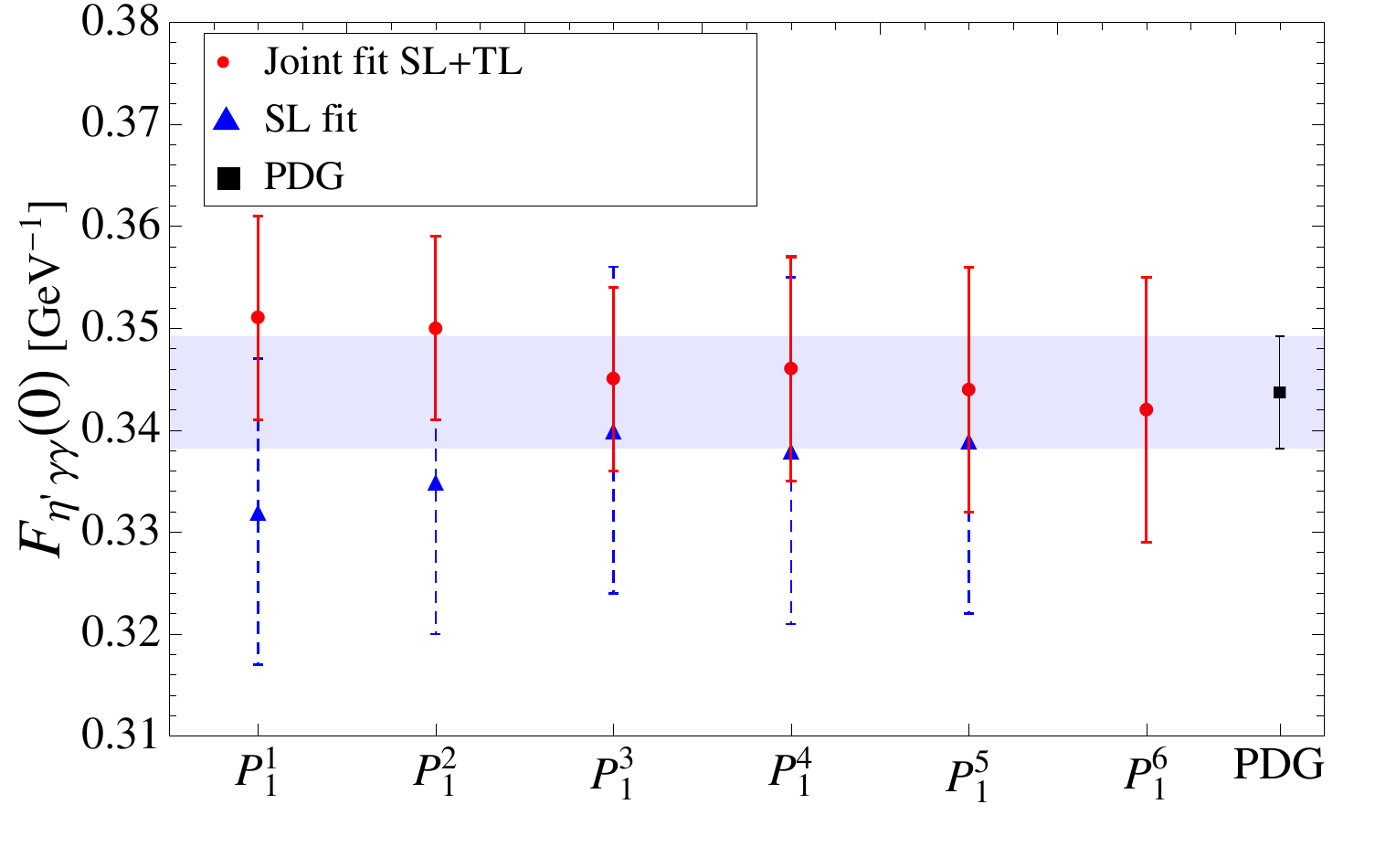}
\includegraphics[width=\columnwidth]{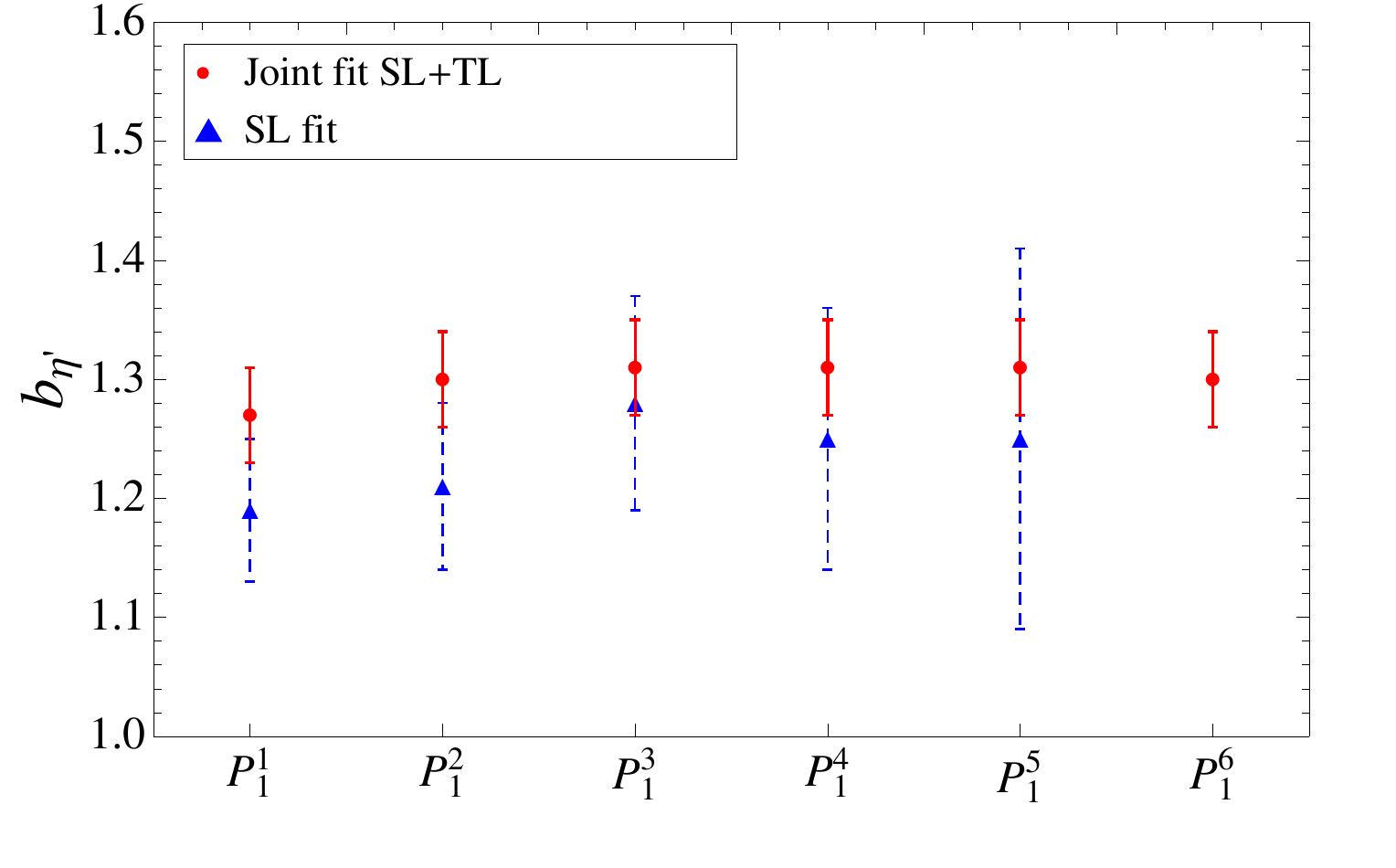}\\[1ex]
\includegraphics[width=\columnwidth]{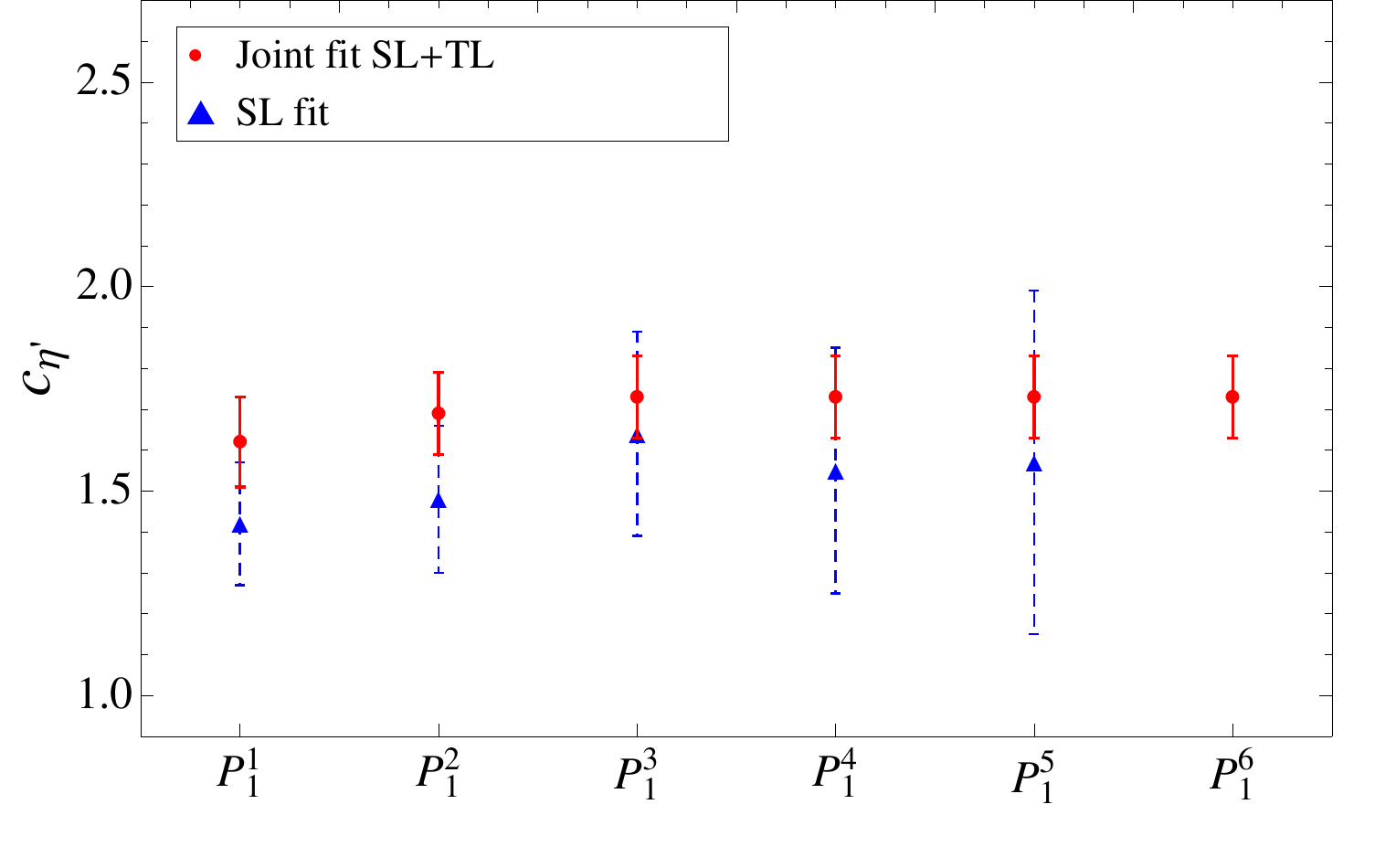}
\includegraphics[width=\columnwidth]{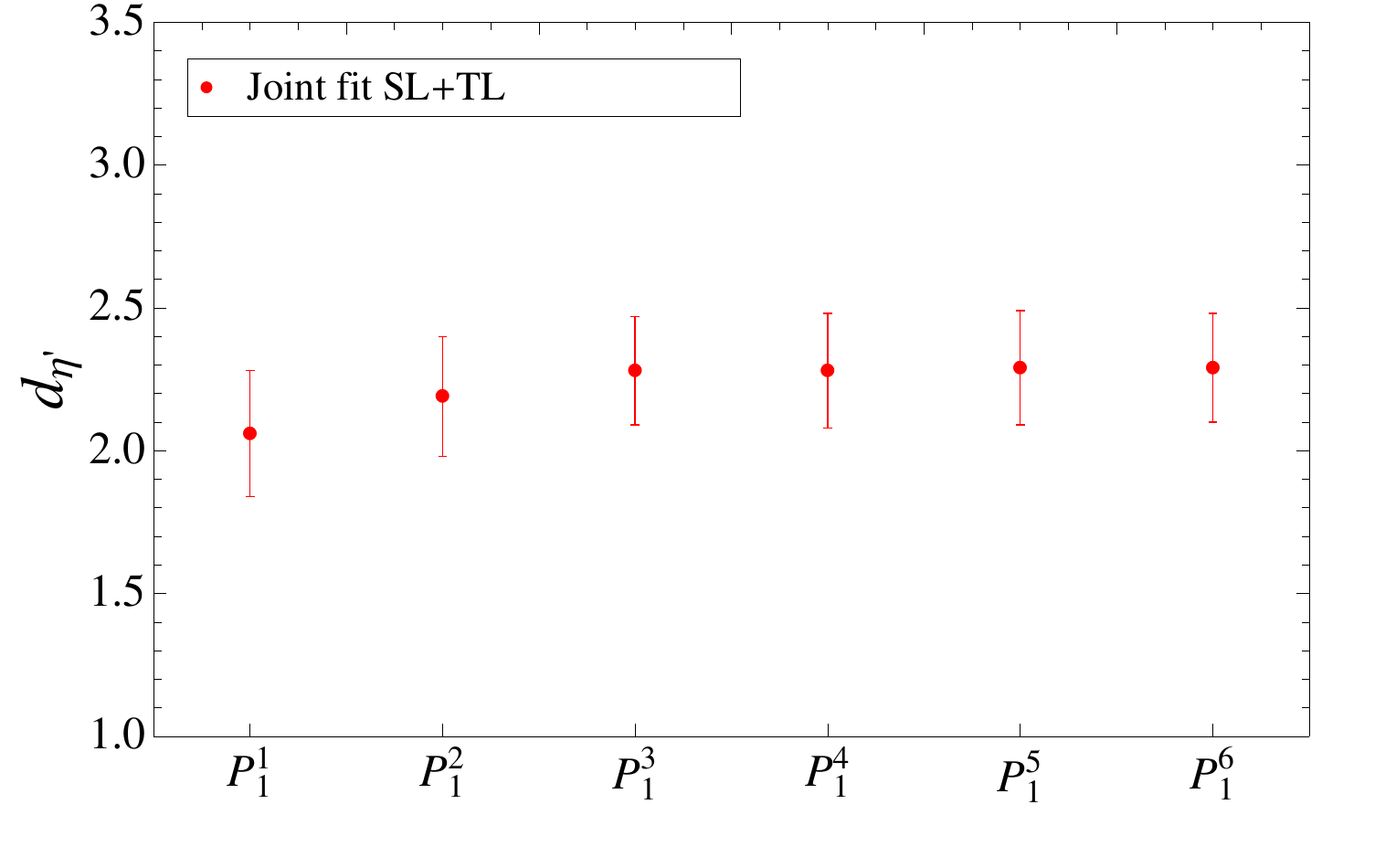}
\caption{Convergence pattern as in Fig.~\ref{Fitpredictionscons} without including information on the
$\Gamma_{\eta^\prime\to\gamma\gamma}$.}
\label{Fitpredictions}
\end{figure*}

We start fitting with a Pad\'{e} approximants' sequence of the type $P_{1}^{L}(Q^{2})$, and current data allow us to reach $L=7$.
We provide a graphical account of our fits as compared to both SL and TL in Figure \ref{Fitstodata},
from where one can see that the one sigma error band associated to the time-like $\eta^{\prime}$ TFF
has considerably decreased as compared to Figure \ref{Fig1}.
The LEPs obtained from the fit are collected in Table~\ref{JointFit} and their corresponding convergence pattern in
Figures \ref{Fitpredictionscons} and \ref{Fitpredictions} (red circles) reflect the impact of the inclusion of TL compared with the old results.
In the table we also provide the pole of the PA.
The coefficients of our best fit are gathered in Appendix~\ref{AppTL}.

Comments on these results are in order:
\begin{enumerate}
\item
The precision gained on the LEPs determination is remarkable as compared to our previous results (blue triangles)
when only SL were fitted~\cite{Escribano:2013kba};
\item
We enlarge our PA sequence by one element (reducing then the systematic uncertainty);
\item
The new LEPs sequence reaches faster the stability value manifesting the excellent performance of the method
as new experimental data is included;
\item
Including $F^{{\rm{\rm exp}}}_{\eta^{\prime}\gamma\gamma}(0)$ as an additional datum in the fit reduces significantly
the uncertainty associated to this quantity.
Regarding to this constraint,
it is noticed that while LEPs obtained from the $P_{1}^{L}(Q^{2})$ sequence are basically insensitive to this effect,
the LEPs obtained from the $P_{1}^{1}(Q^{2})$ element are not and suffer small distortions. 
\end{enumerate}

After the first combined analysis of both SL and TL data,
our central value results for $F_{\eta^{\prime}\gamma\gamma}(0)$ and LEPs are 
\begin{equation}
\begin{array}{l}
F_{\eta^{\prime}\gamma\gamma}(0)=0.344(5)(0)\ {\rm{GeV}}^{-1}\ ,\quad b_{\eta^{\prime}}=1.31(4)(1)\ ,\\[1ex]
c_{\eta^{\prime}}=1.74(9)(3)\ ,\quad d_{\eta^{\prime}}=2.30(19)(21)\ ,
\end{array}
\label{centralresults}
\end{equation}
where the first error is statistic and the second systematic,
the latter being $0\%$ for the value at the origin,
and $1\%, 2\%$, and $9\%$ for the slope, curvature, and third derivative, respectively~\cite{Escribano:2013kba}.
The systematic error can be evaluated as well from the difference between two consecutive elements in the PA sequence~\cite{Baker,Masjuan:2009wy}. However, as illustrated in Figs.~\ref{Fitpredictionscons} and~\ref{Fitpredictions}, this difference is basically negligible for the value at the origin and all the derivatives shown, and we prefer to consider the larger systematic errors reported in Ref.~\cite{Escribano:2013kba} which were obtained after applying the PA method to a set of selected models. The results above can be compared with the ones obtained by the $P^N_N(Q^2)$ sequence in Table~\ref{JointFit}. Since this second sequence stops at its first element (which is actually the first element on the $P^L_1(Q^2)$ sequence as well), we do not consider its results for a combined weighted average.

The systematic error is at the level of the statistical one.
To reduce it, we would need more precise high-energy data on the one hand, and enlarge, on the other hand,
the $P^N_N(Q^2)$ sequence which is limited in this analysis to its first element.
Notice that the $P^N_N(Q^2)$ has systematic errors dramatically smaller than the ones considered here
(see the Appendix in~\cite{Escribano:2015nra} for details).
It turns out that the $\eta^\prime$ TFF is very much dominated by a single hadronic scale that gives to the TFF
its characteristic vector meson dominance-like shape (VMD).
A $P^2_2(Q^2)$ fit cannot be accommodated at the current level,
and we hope that more data from BESIII, MAMI, and Belle-II will help to improve the present values (see the discussion at the end of this section). 
These results can be compared with $F_{\eta^{\prime}\gamma\gamma}(0)=0.344(4)(0)$ GeV$^{-1}$,
$b_{\eta^{\prime}}=1.30(15)(7)$ and $c_{\eta^{\prime}}=1.72(47)(34)$,
obtained using SL data only~\cite{Escribano:2013kba}.
Clearly, the statistical uncertainty of the LEPs has considerably diminished as a consequence of including TL data to the analysis,
being that one of the main results of this work.
Our slope, $b_{\eta^{\prime}}=1.31(4)$, can be compared with the values 1.46(23), 1.24(8) and 1.6(4),
quoted by the CELLO \cite{Behrend:1990sr}, CLEO \cite{Gronberg:1997fj} and Lepton-G (cited in~\cite{Landsberg:1986fd}),
respectively.
One should notice that all the previous collaborations used a single-pole model, VMD, to extract the slope,
which is nothing but the simplest $P_{1}^{1}(Q^{2})$ element from our approach (which we neglected).
Other theoretical predictions existent in the literature are $b_{\eta^{\prime}}=1.47$ predicted by chiral perturbation theory for
$\sin{\theta_{P}}=-1/3$, being $\theta_{P}$ the $\eta$-$\eta^{\prime}$ mixing angle,
$b_{\eta^{\prime}}=1.30$ from constituent-quark loops,
both values taken from~\cite{Ametller:1991jv},
$b_{\eta^{\prime}}=1.33$ from VMD \cite{Bramon:1981sw}, and
$b_{\eta^{\prime}}=2.11$ from the Brodsky-Lepage interpolation formula~\cite{Brodsky:1981rp}.
More recently, one can find $b_{\eta^{\prime}}=1.323(4)$ from resonance chiral theory~\cite{Czyz:2012nq},
$b_{\eta^{\prime}}=1.45^{+0.17}_{-0.12}$ using dispersive techniques \cite{Hanhart:2013vba},
and $b_{\eta^{\prime}}=1.06$ or 1.16 from anomaly sum rules~\cite{Klopot:2013laa}.

The main difference between Figure \ref{Fig1} and Figure \ref{Fitstodata}, left panel, is the width of the uncertainty band,
specially at large $\sqrt{s}$, which is the region where we expect the PA to eventually fail.
To control on the quality of the fits at this large $\sqrt{s}$,
we have repeated the fits by first enlarging artificially the errors of the last energy points and secondly eliminating subsequently
the last data points.
We have observed a completely stable fit even under these manipulations which only slightly enlarge the slope error but
always keeping the same $\chi^{2}_{\rm dof}$ (degrees of freedom).
We conclude, then, that our final results in (\ref{centralresults})
are robust enough and independent of an eventual failure of the PA method at the highest TL energy point.

We benefit from our results $F_{\eta^{\prime}\gamma\gamma}(0)=0.344(5)$ GeV$^{-1}$ and
$F_{\eta^{\prime}\gamma\gamma}(0)=0.342(13)$ GeV$^{-1}$ (constrained and unconstrained cases, respectively)
to predict the $\eta^{\prime}$ partial decay width to two photons.
For the constrained fit, i.e. including the value at the origin in our data set,
the fit returns $\Gamma_{\eta^{\prime}\to\gamma\gamma}=4.35(13)$~keV,
slightly better than the PDG fitted value and at $0.3$ standard deviations off its averaged result.
For the unconstrained case, we find $\Gamma_{\eta^{\prime}\to\gamma\gamma}=4.30(33)$ keV,
which lies $0.1$ standard deviations off the experimental value. 
Regarding the asymptotic behavior of the TFF,
we have considered the $P_{N}^{N}(Q^{2})$ sequence since they have the right asymptotic fall-off $1/Q^{2}$ built-in.
We reached $N=1$ and then predicted the leading coefficient
\begin{equation}\label{etapasym}
\lim_{Q^{2}\to\infty}Q^{2}F_{\eta^{\prime}\gamma^{*}\gamma}(Q^{2})=0.254(3)\ {\rm{GeV}}\ ,
\end{equation}
which is in very good agreement with the value $0.254(21)$ GeV\footnote{Such
value is obtained from the BABAR result $0.251(19)(8)$ GeV after taking into account
kinematical corrections~\cite{Escribano:2013kba}.}
measured at $q^{2}=-112$ GeV$^{2}$ by the BABAR collaboration \cite{Aubert:2006cy} and assuming that the space- and time-like asymptotic duality already holds at that $q^2$.
This prediction is basically the same one obtained in~\cite{Escribano:2013kba} when only the SL data were considered.
Therefore, the effect of including the TL data is negligible in this respect.
Ideally, it would be desirable to extract such value from the $N=2$ element,
which allows for diminishing the intrinsic systematic error (not yet evaluated) as well as for checking convergence.
This should be possible in the future if new precise Belle-II data becomes available.

The result in Eq.~(\ref{etapasym}) needs to be upgraded to include a theoretical error, then. Such error can be obtained in the same way as the systematic error for the TFF's LEPs~\cite{Masjuan:2012wy,Escribano:2013kba,Masjuan:2008fv}. This is, by comparing the asymptotic value of the model used in Refs.~\cite{Masjuan:2012wy,Escribano:2015nra} (the holographic confining model defined in Appendix B of~\cite{Escribano:2015nra}, rescaled to yield the same asymptotic value as in Eq.~(\ref{etapasym})), with the expansion at $Q^2 \to \infty$ of the PA fits to pseudodata, we extract the theoretical uncertainty in percentage collected in Table~\ref{PAinf}.

\begin{table}[htp]
\caption{Theoretical error for the first asymptotic coefficient of the $P^N_{N}(Q^2)$ sequence. First line corresponds to the actual coefficient $\lim_{Q^2 \to \infty} P^N_{N}(Q^2)$. Second line collects its relative error with respect to $\lim_{Q^2 \to \infty} Q^2 F_{\eta'\gamma\gamma}(Q^2)$. See the text for details.}
\begin{center}
\begin{tabular}{c||c|c|c|c}
\hline
$Q^2 F_{\eta'\gamma\gamma}(Q^2)$& $P^1_1(Q^2)$ & $P^2_2(Q^2)$ & $P^3_3(Q^2)$ & $P^4_4(Q^2)$\\
\hline
0.254 & 0.348  & 0.247 & 0.254 & 0.254\\
& 27.0 $\%$ & 2.8$\%$ & 0.1$\%$ & 0.0$\%$\\
\hline
\end{tabular}
\end{center}
\label{PAinf}
\end{table}%

This Table~\ref{PAinf} shows that the theoretical uncertainty associated to the extraction of the TFF's asymptotic value using the simplest  $P^1_1(Q^2)$ is of about $25\%$. This result seems to disagree with the fit shown in Fig.~\ref{Fitstodata}, left panel, where the interpolation of the  $P^1_1(Q^2)$ at around $Q^2 = 20$-$35$ GeV$^2$ is very good, much better than a $25\%$ (the error at $Q^2 \to \infty$ and at $Q^2 = 20$-$35$ GeV$^2$ is basically the same). This disagreement comes from a peculiarity of the $\eta'$-TFF. We have already discussed our impossibility of fitting with a $P^2_2(Q^2)$. This is not an anecdote since it is telling us that a $P^1_1(Q^2)$ is already an excellent description of the TFF. 

Since the difference $P^1_1(Q^2) - P^2_2(Q^2)$ is basically zero, their difference at $Q^2 \to \infty$ is also zero, which implies that the systematic error we have by extracting the asymptotic value with the $P^1_1(Q^2)$ is basically the same as if we would use the $P^2_2(Q^2)$. To be a bit more specific, if we consider the asymptotic value obtained with our fits using the $P^1_1(Q^2)$ and reconstruct the $P^2_2(Q^2)$ using the LEPs obtained with the fit of the $P^7_1(Q^2)$, the relative error between both predictions amounts to a $4\%$. From Table~\ref{PAinf}, the $P^2_2(Q^2)$ induces a theoretical error for the asymptotic coefficient of a $3\%$. We can then conclude that in our case of study, the theoretical error induced by using the $P^1_1(Q^2)$ to extract that asymptotic coefficient is the combination in quadrature of both sources of error, i.e., a final $5\%$, much smaller than the generic $25\%$ quoted in Table~\ref{PAinf}. Let us remark that this $5\%$ is valid only for the present case of the $\eta'$ TFF. 

We should also include a theoretical error to the asymptotic value of the $\eta$ TFF of about $3\%$, which combined with the statistical error obtained in Ref.~\cite{Escribano:2015nra} turns out to read $0.177(16)$GeV.

\section{A reassessment of the $\eta$-$\eta^\prime$ mixing}
\label{mixing}

In this section we reanalyze the $\eta$-$\eta^\prime$ mixing as we did in \cite{Escribano:2013kba,Escribano:2015nra},
and consider the so-called octet-singlet basis, where the $\eta$ and $\eta^\prime$ pseudoscalar decay constants are
defined in terms of the axial currents $J_{5\mu}^a=\overline{q}\gamma_{\mu}\gamma_5\frac{\lambda^a}{\sqrt{2}}q$
as $\langle{0}| J_{5\mu}^a |{P}\rangle=i\sqrt{2}F_P^a p_{\mu}$,
where $a=(0,8)$ refers to its singlet and octet components, respectively. 
The decay constants in terms of the two angles $\theta_0$ and $\theta_8$ read 
\begin{equation}
(F_P^{80})\equiv\left(
\begin{array}{cc}
F_{\eta}^{8}&F_{\eta}^{0}\\[0.5ex]
F_{\eta^{\prime}}^{8}&F_{\eta^{\prime}}^{0}
\end{array}
\right)
=
\left(
\begin{array}{cc}
F_{8}\cos\theta_{8}&-F_{0}\sin\theta_{0}\\[0.5ex]
F_{8}\sin\theta_{8}&F_{0}\cos\theta_{0}
\end{array}
\right)\ .
\label{eq:decayconstantsmatrix}
\end{equation}
In this basis, large-$N_c$ ChPT at NLO predicts \cite{Escribano:2005qq,Feldmann:1999uf}
\begin{eqnarray}
\label{eq:chptF8F0}
&&F_8^2 = \frac{4F_K^2 - F_{\pi}^2}{3}\ , \quad
F_0^2 = \frac{2F_K^2 + F_{\pi}^2}{3} + F_{\pi}^2\Lambda_1\ ,\\
\label{eq:chptangle}
&&F_8 F_0\sin(\theta_8 - \theta_0) = -\frac{2\sqrt{2}}{3}(F_K^2 - F_{\pi}^2)\ ,
\end{eqnarray}
where $F_K\simeq 1.20 F_\pi$ is the kaon decay constant.

At this point we call the attention that $F_0$ is renormalization group (RG) dependent ($F_0=F_0(\mu)$).
This is connected to the $J_{5\mu}^0$ anomalous dimension implying~\cite{Leutwyler:1997yr,Agaev:2014wna}
\begin{equation}
\mu \frac{dF_0}{d\mu} = -N_F\left( \frac{\alpha_s(\mu)}{\pi} \right)^2 F_0+{\cal O}(\alpha_s^3)\ ,
\end{equation}
where $N_F$ is the number of active flavors at the scale $\mu$.
Solving this equation up to order $\alpha_s(\mu)$, the singlet decay constant at a different scale can be expressed as 
\begin{equation}
\begin{array}{rcl}
F_0(\mu) &=&
F_0(\mu_0)\left[1+\frac{2N_F}{\beta_0}\left( \frac{\alpha_s(\mu)}{\pi} - \frac{\alpha_s(\mu_0)}{\pi} \right)\right]\\[1.5ex]
&\equiv& F_0(1+\delta)\ ,
\end{array}
\end{equation}
with $\beta_0=11-2N_F/3$. The parameters $\delta$ and $\Lambda_1$ are interrelated since~\cite{Leutwyler:1997yr}
\begin{equation}\label{l1delta}
\mu \frac{d}{d\mu}\frac{F_0}{\sqrt{1+\Lambda_1}} = 0\ ,
\end{equation}
at NLO in large-$N_c$ ChPT. This equation also implies that if $\Lambda_1=0$, then $\delta=0$.

In the octet-singlet basis, the different limiting behaviors of the TFF, $F_{P\gamma\gamma}\equiv F_{P\gamma^*\gamma}(0)$
and $P_{\infty} \equiv \lim_{Q^2\to\infty}Q^2F_{P\gamma^*\gamma}(Q^2)$,
take the simple form
\begin{eqnarray}
&&F_{\eta\gamma\gamma} = \frac{1}{4\pi^2}\frac{\hat{c}_8 F_{\eta^\prime}^0 - \hat{c}_0
F_{\eta^\prime}^8}{F_{\eta^\prime}^0 F_{\eta}^8 - F_{\eta^\prime}^8 F_{\eta}^0}\ ,
\label{eq:Feta0}\\[1ex] 
&&F_{\eta^\prime\gamma\gamma} = \frac{1}{4\pi^2}\frac{\hat{c}_8 F_{\eta}^0 - \hat{c}_0
F_{\eta}^8}{F_{\eta}^0 F_{\eta^\prime}^8 - F_{\eta}^8 F_{\eta^\prime}^0}\ ,
\label{eq:Fetap0}\\[1ex]
&&\eta_{\infty} =  2(\hat{c}_8 F_{\eta}^8 + \hat{c}_0(1+\delta_\infty) F_{\eta}^0)\ ,
\label{eq:Infeta}\\[1ex]
&&\eta^\prime_{\infty} = 2(\hat{c}_8 F_{\eta^\prime}^8 + \hat{c}_0(1+\delta_\infty) F_{\eta^\prime}^0)\ ,
\label{eq:Infetap}
\end{eqnarray}
where $\hat{c}_8$ and $\hat{c}_0$ are charge factors and
$\delta_\infty=-0.10(1)$~\cite{Escribano:2015nra} accounts for the $F_0$ running from $\mu_0=1$ GeV up to
$\mu\rightarrow\infty$~\cite{Agaev:2014wna}. The error quoted for the $\delta_\infty$ parameter comes from the uncertainty of $\alpha_s(M_z)=0.1182(16)$~\cite{Agashe:2014kda}.

In addition, we want to include in the previous set of equations the OZI-rule--violating parameter $\Lambda_3$, 
neglected in our previous studies, since it is the first correction to the $F_{P\gamma\gamma}$ in large-$N_c$ ChPT, even though belongs formally to the NNLO order. To be consistent with the counting, if we include OZI-rule--violating parameters, we should also take into account mass corrections to the pseudoscalar-into-two-photons decay widths. Such corrections can be directly calculated from the corresponding Lagrangian~\cite{Leutwyler:1997yr}:

\begin{equation}\label{L}
{\cal L}_{P\gamma\gamma} = - \frac{\alpha N_c}{4\pi}\left( \langle Q^2 \phi  \rangle + \frac{\Lambda_3}{3}\langle Q^2 \rangle \langle \phi \rangle+ K_2 \langle Q^2 \chi \phi  \rangle \right) F_{\mu\nu} \tilde{F}^{\mu\nu}\, ,
\end{equation}
\noindent
with $\chi = 2 B M$, $M=diag(\hat{m},\hat{m},m_s)$, $\hat{m}=\frac12 (m_u+m_d)$, $Q^2$ the electric charge of the quarks and $\phi$ the $3\times 3$ matrix representing the nine pseudoscalar fields $\phi^0(x), \dots, \phi^8(x)$.  For $\pi^0 \to\gamma\gamma$:
\begin{eqnarray} \nonumber
F_{\pi^0 \gamma\gamma}(0,0)
&=&
\frac{1}{4 \pi^2 F_{\pi} }(1 + K_2 M_{\pi}^2) \ , \\ \nonumber
\Gamma_{\pi^0 \gamma\gamma}& =& \frac{\alpha_{em}^2 M_{\pi}^3}{64 \pi^3 F_{\pi}^2} (1+ K_2 M_{\pi}^2)^2\, .
\label{Pgammagamma0}
\end{eqnarray}

The value $\Gamma_{\pi^0 \to\gamma\gamma}=(7.63 \pm 0.16)\cdot 10^{-9}$~GeV~\cite{Agashe:2014kda} translates into $K_2 = -0.45 \pm 0.57$, compatible with zero but with a large central value. This suggests that a better experimental resolution for $\pi^0 \to \gamma \gamma$ will have an impact in the $\eta$-$\eta'$ mixing even neglecting isospin corrections since Eq.~(\ref{L}) implies
\begin{eqnarray} \nonumber
\hat{c}_8 &=& \frac{1}{\sqrt{3}}\left(1+ \frac13 K_2(7 M_{\pi}^2-4 M_K^2)\right) \\ \nonumber
\hat{c}_0 &=&  \sqrt{\frac83}\left(1+\Lambda_3 + \frac13 K_2(2 M_{\pi}^2 +M_K^2)\right)\, .
\end{eqnarray}

The set (\ref{eq:Feta0}, \ref{eq:Fetap0}, \ref{eq:Infeta}, \ref{eq:Infetap}) form a system of 4 equations with 5 unknowns ($F_{\eta^{(\prime)}}^{(8,0)}, \Lambda_3$).
Then it could seem that, at least taking $\Lambda_3=0$, we may solve the system.
However, as explained in~\cite{Escribano:2013kba}, such a system is underdetermined as the relation

\begin{eqnarray}
\label{eq:deg}
\eta_{\infty}F_{\eta\gamma\gamma}& +& \eta^\prime_{\infty}F_{\eta^\prime\gamma\gamma}= \frac{3}{2\pi^2}\left( 1 + \frac{8}{9}(\delta_\infty +\Lambda_3+\delta_\infty\Lambda_3) \right) \\ \nonumber
&+&
\frac{3}{2\pi^2} \frac{K_2}{27}\left(4 M_K^2(1+2 \delta_\infty) + M_{\pi}^2(23 + 16 \delta_\infty) \right) \, ,
\end{eqnarray}
is free of mixing parameters. Indeed, (\ref{eq:deg}) fixes $\Lambda_3$ once its left-hand side is (experimentally) known.
However, we still have to face the fact that our system is underdetermined.
In order to overcome this problem, we notice that at NLO in large-$N_c$ ChPT, Eqs.~(\ref{eq:chptF8F0},\ref{eq:chptangle}) provide a clean prediction for both $F_8$ and $(\theta_8 - \theta_0)$ in terms of the well-known value for $F_K/F_{\pi}$~\cite{Agashe:2014kda}.
Taking either Eq.~(\ref{eq:chptF8F0}) or Eq.~(\ref{eq:chptangle}) as a constraint, one would add an additional equation to the previous system, which would provide a unique solution. Taking both, would lead to an overdetermined system, which in general has no solution.
For this reason, we adopt a democratic procedure in which we perform a $\chi^2$ fit including both,
Eq.~(\ref{eq:chptF8F0}) and Eq.~(\ref{eq:chptangle}) constraints, together with (\ref{eq:Feta0}, \ref{eq:Fetap0},\ref{eq:Infeta}, \ref{eq:Infetap},\ref{eq:deg}).

Using the following parameters as inputs for the $\chi^2$ function
\begin{eqnarray} \nonumber
F_{\eta\gamma\gamma}^{\rm exp} &=& 0.2738(47)~\textrm{GeV}^{-1}\, , \quad F_{\eta^{\prime}\gamma\gamma}^{\rm exp} = 0.3437(55)~\textrm{GeV}^{-1}\\ \nonumber
 \eta_{\infty}^{\rm exp} &=& 0.177(16)~\textrm{GeV}\, , \hspace{0.9cm}  \eta_{\infty}^{\prime exp} = 0.254(13)~\textrm{GeV} \\ \nonumber 
 F_K/F_{\pi} &=& 1.198(5) \\ \nonumber
 K_2 &=& -0.45(57)\, , \hspace{1.9cm} \delta_{\infty} = -0.10(1)\, , \nonumber
\end{eqnarray}
we would obtain the mixing parameters collected in Table~\ref{tab:81res}, first column. The total $\chi^2=3.45$ has a p-value=0.18 which is acceptable with two degrees of freedom.

By observing the different terms contributing to the $\chi^2$ function, we realize that the piece related with the determination of the $\eta$ TFF at $Q^2 \to \infty$ is the one most disfavored (contributing with $1.97$ to the $\chi^2$), with a theoretical prediction of $0.155(19)$GeV to be compared with the experimental counterpart of $0.177(16)$GeV. Such a discrepancy of about $0.9\sigma$ can be attributed to a rather small value of $\theta_0$. The smallness of $\theta_0$ comes ultimately from the experimental value of $\eta^{\prime exp}_{\infty}$. This is so because in this observable, the angle $\theta_0$ comes with $\cos\theta_0\sim 1$ and then, $\eta^{\prime exp}_{\infty}$ determines the value of $F_0$ which, in turn, and through Eq.~(\ref{eq:deg}) determines $\theta_0$. To test this correspondence, we could imagine a $\eta^{\prime exp}_{\infty}$ $10\%$ higher than what we found from the fits. This new value would be in agreement with the BABAR measurement at $q^2=-112$ GeV$^2$, transferred to the space-like region including a $10\%$ effect on the violation of the space- and time-like duality as suggested in Ref.~\cite{Escribano:2015nra}. With such enhancement, the value of $F_0$ will be as well proportionally enhanced. Then, the fit will return $\theta_0$ a $10\%$ larger, a $\chi^2$ result of about $1.9$, and a prediction for $\eta_{\infty}$ $5\%$ larger. As we argued before, experimental data for the $\eta^{\prime}$ TFF at large momentum transfer comes exclusively from BABAR collaboration and a second set of experimental data would be highly welcome to settle this issues.

Alternatively, we can ascribe the smallness of the $\eta_{\infty}$ to an underestimation of a theoretical error without the need to \textit{touch} any number coming from experimental determination. Up to now, we have assumed that the expressions for $F_{0,8}$ and $\theta_{0,8}$ calculated at NLO in large-$N_c$ ChPT contain no theoretical error coming from neglecting higher orders in that expansion. This is an hypothesis that can be tested by allowing higher order effects in the definition of $F_K/F_{\pi}$. We assumed that the departure of that ratio from $1$ came from the NLO alone. We can extend this argument by saying that $F_K/F_{\pi} -1 = \epsilon + \epsilon^2 = 0.198$ which implies $\epsilon = 0.17$ and $\epsilon^2 = 0.03$. $\epsilon$ is the small quantity representing the NLO and $\epsilon^2$ the NNLO of about $2.4\%$ of $F_K/F_{\pi}$. This hypothesis is confirmed both from $SU(3)$ ChPT~\cite{Bijnens:2014lea} and  from $SU(3)$ ChPT in the large-$N_c$ limit~\cite{Ecker:2013pba}. Whit such new error for $F_K/F_{\pi}$ we can repeat the fit to obtain the mixing parameters. The new $\chi^2 = 2.01$ represents a p-value = 0.37 and $\chi^2_{\rm{dof}}=1.00$. The results of the mixing are collected in the second row of Table~\ref{tab:81res} and is the main result of this section.
The net effect of including an extra source of theoretical error on $F_K/F_{\pi}$ turns out to be a larger $\theta_0$ by a $25\%$ which in turn comes from a better prediction of $\eta_{\infty}$, which now reads $0.164(33)$GeV.
The $\chi^{2}_{\rm dof}$ value is excellent,
which indicates a good agreement with large-$N_c$ ChPT but with non-negligible NNLO corrections.
In addition, we can use (\ref{eq:chptF8F0}) to predict the value $\Lambda_1=0.01(13)$.
\begin{table}
\centering
\caption{Predictions for the mixing parameters. $\theta_{8,0}$ are expressed in degrees. First row is our basic scenario while second row (our preferred scenario) includes, on top, a theoretical error for $F_K/F_{\pi}$. Third row corresponds to setting $K_2 = 0$ and $\delta_{\infty} = 0$ in the fit function. See the main text for details.}
\begin{tabular}{ccccccc}
\hline
$F_8/F_\pi$ & $F_0/F_\pi$ & $\theta_8$ & $\theta_0$ & $\Lambda_3$  & $\Lambda_1$ & $\chi^2_{\rm{dof}}$\\
\hline
$1.26(1)$ & $1.15(5)$ & $-21.9(1.7)$ & $-5.6(2.0)$ & $-0.04(7)$ & $0.04(13)$ &$1.73$\\
 ${\bf 1.27(2)}$ & ${\bf 1.14(5)}$ & ${\bf -21.2(1.9)}$ & ${\bf -6.9(2.4)}$ & ${\bf -0.02(7)}$ & ${\bf 0.01(13)}$ &${\bf 1.00}$\\
\hline
$1.28(2)$ & $1.04(3)$ & $-22.3(0.8)$ & $-7.1(2.0)$ & $-0.13(3)$ & $-0.21(9)$ &$1.04$\\
\hline
\end{tabular}
\label{tab:81res}
\end{table}

The outcome of our fit is teaching us that $\Lambda_{1,3}$ are compatible with zero and one is tempted to eliminate them from our system of equations. This is actually the avenue followed by the FKS scheme~\cite{Feldmann:1998vh} (i.e., $\Lambda_{1,3}=0, K_2=0, \delta_{\infty}=0$) and at first sight it may seem we are confirming their findings. Notice, however, that $\Lambda_1$ and $\delta_{\infty}$ are interrelated~(\ref{l1delta}) and by repeating the fit imposing $\delta_\infty=0$, $K_2=0$ but $\Lambda_{1,3} \neq0$, our fit returns non-negligible values for $\Lambda_{1,3}$ and a set of mixing parameters, collected in the last row of Table~\ref{tab:81res}, still compatible with our previous findings. Our results represent then an update of the FKS scheme including all NLO corrections together with an estimate of NNLO effects, which cannot be neglected in order to obtain a good $\chi^2$ (cf. first against second rows in Table~\ref{tab:81res}).

\begin{figure*}
\centering
\includegraphics[width=\columnwidth]{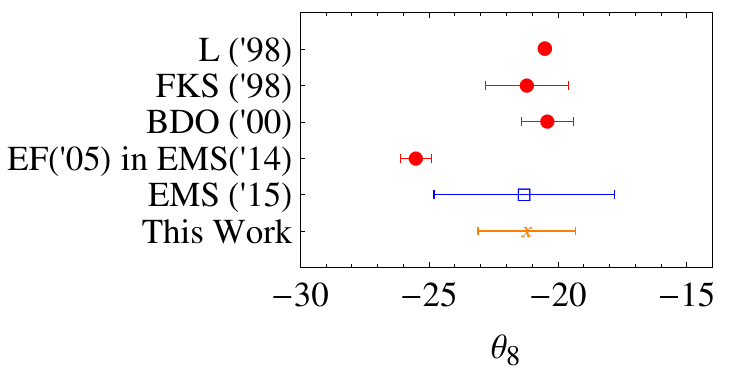}
\hspace{-3.5cm}
\includegraphics[width=\columnwidth]{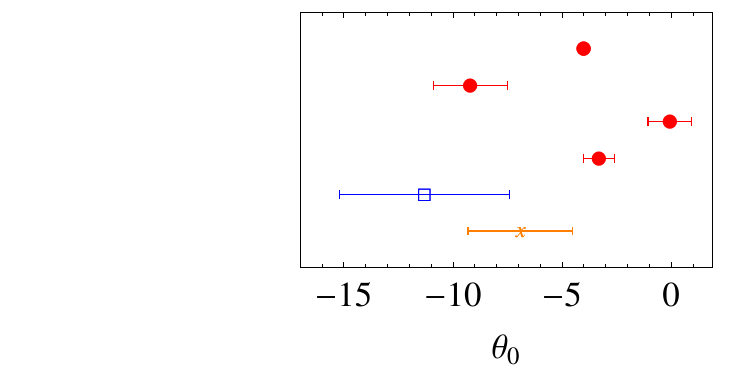}\\
\includegraphics[width=\columnwidth]{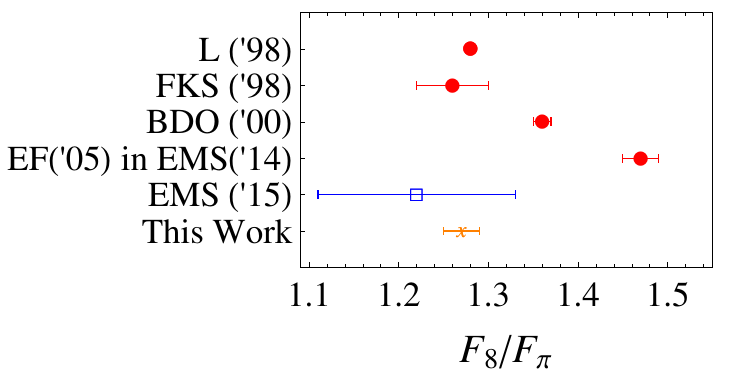}
\hspace{-3.5cm}
\includegraphics[width=\columnwidth]{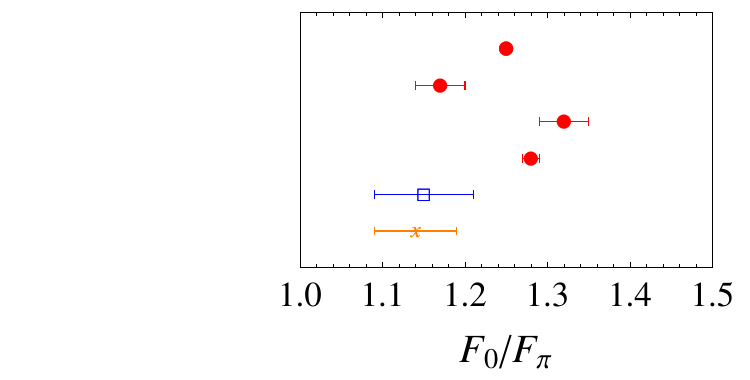}
\caption{$\eta$-$\eta^\prime$ mixing parameters in the octet-singlet basis from
L~\cite{Leutwyler:1997yr}, FKS~\cite{Feldmann:1998vh}, BDO~\cite{Benayoun:1999au},
EF~\cite{Escribano:2005qq}, EMS(14)~\cite{Escribano:2013kba}, and EMS(15)~\cite{Escribano:2015nra}.}
\label{fig:mixing80}
\end{figure*}

In Figure~\ref{fig:mixing80}
we collect our main result (orange crosses) from the second row of Table~\ref{tab:81res} and compare them to different predictions in the
literature~\cite{Leutwyler:1997yr,Feldmann:1998vh,Benayoun:1999au,Escribano:2005qq} (red dots),
as well with our previous results~\cite{Escribano:2015nra} in blue-empty squares.
We see that the main difference with respect to our previous work~\cite{Escribano:2015nra},
where we did not use the $\eta^\prime$ TFF asymptotic value appears in $\theta_0$.
This is to be expected as the inclusion of $\Lambda_{1}$ and $\Lambda_{3}$ affects the singlet part exclusively.
In addition, we have reduced our errors thanks to the constraints from large-$N_c$ ChPT.
Our prediction for $\Lambda_3$ may be compared with the one in~\cite{Benayoun:1999au}, $\Lambda_3=-0.03(2)$. 
Both of them point towards a small value for this parameter, and agree with its sign.
We find that $\Lambda_3$ actually plays an important role not only in fulfilling the degeneracy condition (\ref{eq:deg}),
but in the $\eta(\eta^\prime)\rightarrow\gamma\gamma$ decays as well.
In addition, the $\Lambda_1$ term is rather important and affects specially the $\eta^\prime$ results,
where deviations of order $10\%$ appear if $\Lambda_1$ is omitted.
Finally, we stress that the use of the RG equation for $F_0$ is fundamental,
whereas most of the theoretical and experimental analysis do not account for this effect,
which ---to our best knowledge--- was included for the first time in~\cite{Agaev:2014wna}.
This effect increases $\eta_{\infty}$ and diminishes $\eta^\prime_{\infty}$, bringing in agreement experiment and theory. 

Our results may be translated to the quark-flavor basis where the decay constants are defined as
\begin{equation}
(F_P^{qs})\equiv\left(
\begin{array}{cc}
F_{\eta}^{q} & F_{\eta}^{s}\\[0.5ex]
F_{\eta^{\prime}}^{q} & F_{\eta^{\prime}}^{s}
\end{array}
\right)
=
\left(
\begin{array}{cc}
F_{q}\cos\phi_{q} & -F_{s}\sin\phi_{s}\\[0.5ex]
F_{q}\sin\phi_{q} & F_{s}\cos\phi_{s}
\end{array}
\right)\ .
\end{equation}

Using the rotation matrix~\cite{Feldmann:1998vh}
\begin{equation} \nonumber
U(\theta_{\rm ideal})=\frac{1}{\sqrt{3}}
\left(
\begin{array}{cc}
1 & -\sqrt{2} \\
\sqrt{2} & 1 
\end{array}
\right)\ ,
\end{equation}
\label{eq:rot}
then
\begin{equation} \nonumber
(F_P^{qs}) = (F_P^{80})U(\theta_{\rm ideal})\ .
\end{equation}

From the above equation, together with the values from the second row of Table~\ref{tab:81res},
we obtain 
\begin{equation}
\label{eqsresultflavor}
\begin{array}{ll}
F_q=1.03(4)F_{\pi}\ ,& \quad F_s=1.36(4)F_{\pi}\ ,\\[1ex]
\phi_q=39.6(2.3)^{\circ}\ ,& \quad \phi_s=40.8(1.8)^{\circ}\ .
\end{array}
\end{equation}

In addition, we can predict the ratio
$R_{J/\psi}\equiv\Gamma_{J/\psi \rightarrow\eta^\prime\gamma}/\Gamma_{J/\psi \rightarrow\eta\gamma}$,
which is given in terms of 
$\phi_q$ alone~\cite{Escribano:2005qq} as 
\begin{equation}\label{EqJpsi}
R_{J/\psi} = \tan^2\phi_q \left( \frac{m_{\eta^\prime}}{m_{\eta}} \right)^4
\left( \frac{M_{J/\psi}^2 - m_{\eta^\prime}^2}{M_{J/\psi}^2 - m_{\eta}^2} \right)^3\ .
\end{equation}
With~(\ref{eqsresultflavor}), $R_{J/\psi}=5.2(9)$, just at $0.5\sigma$ from the experimental value
$R_{J/\psi}=4.7(2)$~\cite{Agashe:2014kda}.
It may be that, as precision improves, the deviation grows, which would be a hint of novel phenomena in the $\eta$-$\eta^\prime$ system,
as gluonium component, which has long been debated, but not found so far~\cite{Escribano:2007cd}.
We recall in this sense that large-$N_c$ ChPT implicitly assumes that such component is not present in the $\eta^\prime$.
Moreover, the 3-gluon annihilation amplitude, not included in our framework, may need to be included to account for this
$10\%$ discrepancy~\cite{Gerard:2013gya}.
Alternatively, we could use  the experimental $R_{J/\psi}$ together with Eq.~(\ref{EqJpsi}) to obtain $\phi_q=38.2(6)^{\circ}$ in agreement with our fit determination. With respect to the $VP\gamma$ couplings calculated in our previous work~\cite{Escribano:2015nra},
the new results yield more precise errors and very similar central values, with the exception of the $\phi$ cases,
which get slightly closer to the experimental results.

With the set of parameters in Table~\ref{tab:81res}, together with (\ref{eq:decayconstantsmatrix}),
we can also predict the ratio $R_{Z}\equiv\Gamma_{Z \rightarrow\eta^\prime\gamma}/\Gamma_{Z \rightarrow\eta\gamma}$,
which is given by~\cite{Feldmann:1998sh}
\begin{equation}
\label{RZ}
R_{Z} = \left| \frac{F_{\eta^{\prime} \gamma Z}}{F_{\eta \gamma Z}}\right|^2
\left( \frac{M_{Z}^2 - m_{\eta^\prime}^2}{M_{Z}^2 - m_{\eta}^2} \right)^3\ ,
\end{equation}
where, assuming the asymptotic behavior, $M_Z^2 F_{P \gamma Z}(M_Z^2)=6\sqrt{2}( C_{8\gamma Z} F_P^8 + C_{1 \gamma Z} F_P^0(1+\delta_\infty))$ with $C_{8\gamma Z} = (1- 4 \sin^2\theta_W)/6\sqrt{6}$, $C_{1\gamma Z} = (2- 4 \sin^2\theta_W)/3\sqrt{3}$ and 
$\theta_W$ the Weinberg angle at $M_Z^2$~\cite{Agashe:2014kda}.
Since $C_{8\gamma Z}\ll C_{1\gamma Z}$, one may expect $R_Z \simeq \cot^2\theta_0 \sim 68(1)$~\cite{Feldmann:1998sh},
an observable quite sensitive, then, to the singlet angle.
However, since $F_{\eta}^8 \gg F_{\eta}^0$,
the denominator of (\ref{RZ}) should not be approximated and all the terms should be retained.
In this respect, we find $R_Z = 11(3)$, indicating still a large singlet component in $R_Z$.

To close this section, we comment on possible venues to improve our errors.
On the one hand, it would be desirable to improve not only on $\eta_{\infty}$,
which now is the input with the largest error, but also to obtain $\eta^\prime_{\infty}$ from a $P^2_2(Q^2)$,
which would reassess both the central value and the error of this parameter.
This would be possible from future Belle-II data. Curiously enough, the $\Gamma_{\pi^0 \to \gamma \gamma}$ measurement is important to study NLO effects and the role of SU(3) breaking in the mixing scheme. 
On the other hand, it would be interesting to have a more precise $\mathcal{O}(\alpha_s)^2$ calculation for $\delta_\infty$ although its impact may be marginal.
A NNLO predictions for the mixing parameters and $\eta(\eta^\prime)\rightarrow\gamma\gamma$ decays will allow to check the stability and accuracy of the results. This calculation will involve new low-energy constants which knowledge is scarce, though.
In addition, future lattice analysis may play an important role in this field~\cite{Michael:2013gka}
and a combined analysis using the PA method will be highly desirable.

\section{Conclusions}
\label{conclusions}

In this work we have shown the excellent performance of the Pad\'e approximants' method
developed in \cite{Masjuan:2012wy,Escribano:2013kba,Sanchez-Puertas:2014zla,Escribano:2015nra}
for the description of the recently reported first observation of the Dalitz decay $\eta^\prime\to\gamma e^+e^-$
by the BESIII Collaboration \cite{Ablikim:2015wnx}.
This experimental analysis studies the time-like region of the $\eta^\prime$ transition form factor up to the resonance region.

Unlike our previous works, we have explored in the present one the limits of application of PAs in the TL region
finding that, beyond expectations, PAs can be extended to energies very close to the location of poles.
We have nicely described the behavior of the modulus square of the $\eta^\prime$ TFF thus showing that this form factor has
a simple analytical structure in the complex plane made of an isolated branch cut due to the $\pi\pi$ production threshold and a set of single poles.

The careful analysis of the PA sequence $P^L_1(Q^2)$
reveals, however, more effects than those of the $\rho$ resonance emerging here from $\pi\pi$ rescattering.
Subleading effects caused by additional branch cuts or the influence of higher resonances' tails are also captured by PAs
and are indeed responsible for the shift of the PA-pole location with respect to the naive projection of the $\rho$ resonant pole
onto the real axis.
Since this shift is not known with precision it is difficult to extract from the PA pole the exact position of the resonance pole.
This limitation of the method, already mentioned at the beginning of this work,
does not prevent the PAs from guiding us about the underlying analytical structure of the TFF.
One can take advantage of this highly non-trivial knowledge to further use the PAs method in other scenarios
such as $B\to\pi$ semileptonic form factors or combine it with dispersion relations to include resonance poles.

A last remark concerns the role of VMD in experimental analyses, now that the meaning of the PA pole on the real axis is understood.
As pointed out in \cite{Masjuan:2008fv}, VMD should be interpreted as a first step in a systematic approximation,
that is, the $P^1_1$ element belonging to a more general and exhaustive $P^N_1$ sequence.
Although it is common to report on such fit for ease of comparison,
the range of application of VMD in the TL region is much shorter than the $P^7_1$ we used here.

In summary, PAs are not only useful for fitting and extrapolating data within the SL region
but also give us information about the analytical structure of the TFF in the TL low-energy region.
On the one hand, dispersion relations with a single $\pi\pi$ elastic cut
for the isovector part of the TFF and a Breit-Wigner model for the isoscalar one \cite{Hanhart:2013vba,Kubis:2015sga} proves
the TFF to be a Stieltjes function for which the PA convergence is guaranteed in the SL and in the TL below the cut. This already ensures an optimal extraction of the LEPs from experimental data with tiny systematic errors.
On the other hand, the modulus of the TFF along the branch cut is also well reproduced thanks to the smoothness of the opening of the
$\pi\pi$ cut, even though the convergence theorems do not guide about the performance in this region.
PAs are also capable of accommodating the SL region high-energy QCD constraints
while still providing accurate predictions of the $\Gamma_{\eta(\eta^\prime)\to\gamma\gamma}$ decay widths.

Moreover, they allowed us to report the most up-to-date results for slope, curvature, and third derivative of the $\eta'$TFF, and to update the $\eta$-$\eta^\prime$ mixing parameters in a mixing scheme compatible with the most general large-$N_c$ ChPT scenario at NLO, thus superseding the values obtained in our previous works and those from FKS~\cite{Feldmann:1998vh}, BDO~\cite{Benayoun:1999au},
and EF~\cite{Escribano:2005qq} schemes. With such results we predicted the $J/\psi$ and $Z\to\eta^{(\prime)}\gamma$ decays.

\appendix
\section{Best Pad\'e approximant fit parameterisation}
\label{AppTL}
In this appendix, we provide the parameterizations of our best $P^L_1(Q^2)$ fit for the $Q^2 F_{\eta^\prime\gamma^*\gamma}(Q^2)$.
Defining $P^L_1(Q^2)$ as
\begin{equation}\label{PL1}
P^L_1(Q^2) = \frac{T_N(Q^2)}{R_1(Q^2)} = \frac{t_1 Q^2+ t_2 Q^4+\cdots +t_N (Q^2)^N}{1+r_1 Q^2}\ ,
\end{equation}
the corresponding fitted coefficients\footnote{For
full precision of the coefficients together with the correlation matrix, contact the corresponding authors.}
for the $Q^2 F_{\eta^\prime\gamma^*\gamma}(Q^2)$ are collected in Table~\ref{tab:param}.
With these coefficients  one can extract the slope of the TFF by expanding~(\ref{PL1}) and normalizing the result as
\begin{equation}
b_{\eta^\prime} =  m_{\eta^\prime}^2 (t_1 \cdot r_1 - t_2)/t_1 = 1.312\ ,
\end{equation}
with $m_{\eta^\prime} = 0.95778$~GeV, to be compared with the second column in Table~\ref{JointFit}.

\begin{table}
\centering
\caption{Fitted coefficients for the best Pad\'e approximant, $P^7_1(Q^2)$, associated to $Q^2 F_{\eta^\prime\gamma^*\gamma}(Q^2)$.}
\begin{tabular}{cc}
\hline
Coefficient & Value\\
\hline
$t_1$ & $0.3437$\\ 
$t_2$ & $3.847\cdot 10^{-3}$\\
$t_3$ & $0.550\cdot 10^{-3}$\\
$t_4$ & $-1.621\cdot 10^{-4}$\\
$t_5$ & $1.338\cdot 10^{-5}$\\
$t_6$ & $-4.495\cdot 10^{-7}$\\ 
$t_7$ & $5.261\cdot 10^{-9}$\\  
\hline
$r_1$ & $1.4413$\\
\hline                                           
\end{tabular}
\label{tab:param}
\end{table}

\begin{acknowledgements}
S.~Gonz\`alez-Sol\'is would like to thank the Institut f\"{u}r Kernphysik of the 
Johannes Gutenberg-Universit\"{a}t in Mainz for hospitality during the performance of this work.
We would also like to thank S.~Eidelman for encouragement and discussions.
This work was supported in part (P.~M., P.~S.~P.) by the Deutsche Forschungsgemeinschaft DFG
through the Collaborative Research Center ``The Low-Energy Frontier of the Standard Model" (SFB 1044).
This work was also supported in part (R.~E., S.~G-S) by the FPI scholarship BES-2012-055371 (S.~G-S),
the Ministerio de Ciencia e Innovaci\'on under grant FPA2011-25948,
the Ministerio de Econom\'ia y Competitividad under grants 
CICYT-FEDER-FPA 2014-55613-P and SEV-2012-0234, 
the Secretaria d'Universitats i Recerca del Departament d'Economia i Coneixement 
de la Generalitat de Catalunya under grant 2014 SGR 1450, 
and the Spanish Consolider-Ingenio 2010 Programme CPAN (CSD2007-00042).
\end{acknowledgements}

\end{document}